\documentclass[superscriptaddress,aps,pra,twocolumn,showpacs,nofootinbib,longbibliography,notitlepage,floatfix]{revtex4-2}
\usepackage{amsmath,amssymb,amsthm}
\usepackage[colorlinks=true,citecolor=blue,urlcolor=blue]{hyperref}
\makeatletter
\providecommand \href@noop [1]{#1}
\makeatother
\usepackage[pdftex]{graphicx}
\usepackage{times,txfonts}
\usepackage{braket}
\usepackage{color}
\usepackage{xcolor}
\usepackage{natbib}
\usepackage[normalem]{ulem}
\usepackage{mathtools}
\usepackage{ulem}
\usepackage{latexsym}
\usepackage{tabularx, booktabs}
\usepackage{graphics,epstopdf}
\usepackage{graphicx}
\usepackage{float}
\usepackage{amsfonts}
\usepackage{ragged2e}
\usepackage{blindtext}
\usepackage{siunitx}

\newcommand{\figref}[1]{\figurename~\ref{#1}}

\usepackage{bm}
\usepackage{chapterbib}

\newcommand{\be}{\begin{equation}}
\newcommand{\ee}{\end{equation}}
\newcommand{\ba}{\begin{eqnarray}}
\newcommand{\ea}{\end{eqnarray}}

\newcommand{\tr}{\operatorname{Tr}}

\usepackage{multirow}
\usepackage{appendix}
\usepackage{url}

\begin{document}
%
\title{Demonstration of Hardware Efficient Photonic Variational Quantum Algorithm}

\author{Iris Agresti\textsuperscript{*}}
\affiliation{University of Vienna, Faculty of Physics, Vienna Center for Quantum
Science and Technology (VCQ), Boltzmanngasse 5, Vienna 1090, Austria}

\author{Koushik Paul\textsuperscript{*}}
\affiliation{Department of Physical Chemistry, University of the Basque Country UPV/EHU, Apartado 644, 48080 Bilbao, Spain}
\affiliation{EHU Quantum Center, University of the Basque Country UPV/EHU, Barrio Sarriena, s/n, 48940 Leioa, Spain}

\author{Peter Schiansky\textsuperscript{*}}
\affiliation{University of Vienna, Faculty of Physics \& Research Network Quantum \& Aspects of Space Time (TURIS),\\ Boltzmanngasse 5, 1090 Vienna, Austria}

\author{Simon Steiner}
\affiliation{University of Vienna, Faculty of Physics \& Research Network Quantum \& Aspects of Space Time (TURIS),\\ Boltzmanngasse 5, 1090 Vienna, Austria}

\author{Zhenghao Yin}
\affiliation{University of Vienna, Faculty of Physics, Vienna Center for Quantum
Science and Technology (VCQ), Boltzmanngasse 5, Vienna A-1090, Austria}
\affiliation{University of Vienna, Faculty of Physics, Vienna Doctoral School of Physics (VDSP), Boltzmanngasse 5, Vienna A-1090, Austria}

\author{Ciro Pentangelo}
\affiliation{Istituto di Fotonica e Nanotecnologie, Consiglio Nazionale delle Ricerche (IFN-CNR), piazza L. Da Vinci 32, 20133 Milano, Italy}

\author{Simone Piacentini}
\affiliation{Istituto di Fotonica e Nanotecnologie, Consiglio Nazionale delle Ricerche (IFN-CNR), piazza L. Da Vinci 32, 20133 Milano, Italy}

\author{Andrea Crespi}
\affiliation{Dipartimento di fisica, Politecnico di Milano, piazza L. Da Vinci 32, 20133 Milano, Italy}


\author{Yue Ban}
\affiliation{Departamento de F\'{i}sica, Universidad Carlos III de Madrid, Avda. dela Universidad 30, 28911 Leganes, Spain}

\author{Francesco Ceccarelli}
\affiliation{Istituto di Fotonica e Nanotecnologie, Consiglio Nazionale delle Ricerche (IFN-CNR), piazza L. Da Vinci 32, 20133 Milano, Italy}

\author{Roberto Osellame}
\affiliation{Istituto di Fotonica e Nanotecnologie, Consiglio Nazionale delle Ricerche (IFN-CNR), piazza L. Da Vinci 32, 20133 Milano, Italy}

\author{Xi Chen}
\email{xi.chen@csic.es}
\affiliation{Instituto de Ciencia de Materiales de Madrid (CSIC), Cantoblanco, E-28049 Madrid, Spain}

\author{Philip Walther}
\email{ philip.walther@univie.ac.at}
\affiliation{University of Vienna, Faculty of Physics, Vienna Center for Quantum
Science and Technology (VCQ), Boltzmanngasse 5, Vienna 1090, Austria}
\affiliation{Institute for Quantum Optics and Quantum Information Sciences (IQOQI), Austrian Academy of Sciences, Boltzmanngasse 3, Vienna 1090, Austria}
\affiliation{QUBO Technology GmbH, 1090 Vienna, Austria}

\begin{abstract}
Quantum computing has changed the paradigm of computer science, where quantum technologies have promised to outperform classical ones. Such an advantage was only demonstrated for tasks with no application or out of reach for the state-of-art quantum technologies.
In this context, a promising strategy to find practical use of quantum computers exploits hybrid models, where a quantum device estimates a hard-to-compute quantity, while a classical optimizer trains the parameters.
In this work, we demonstrate that single photons and linear optical networks are sufficient for implementing Variational Quantum Algorithms, when the problem specification (ansatz), is tailored to this specific platform. We show this by a proof-of-principle demonstration to tackle an instance of a factorization task, whose solution is encoded in the ground state of a suitable Hamiltonian. This work which combines Variational Quantum Algorithms with hardware efficient ansatzes for linear-optics showcases a promising pathway towards practical applications for photonic quantum platforms.
\end{abstract}

\maketitle
\section{Introduction}
\label{sec:intro}
In the past decade, quantum computing has risen to the forefront of current technological discussions, fundamentally transforming the field of computation. 
\begin{figure}[tb]
    \centering
    \includegraphics[width=0.45\textwidth]{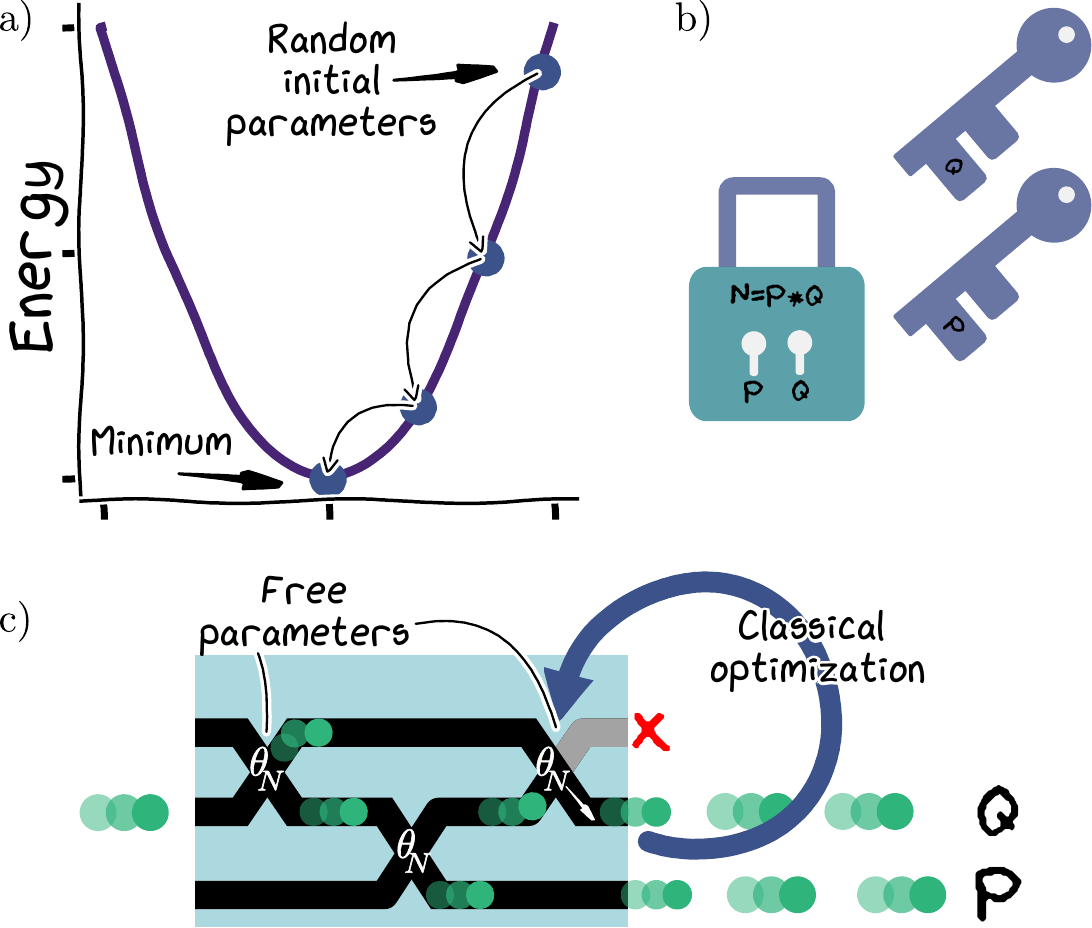}
    \caption{\textbf{Variational Quantum Algorithm to solve a factorization task. a)} In the noisy intermediate scale quantum (NISQ) era, before fully fault-tolerant large-scale quantum computers are available, Variational Quantum Algorithms (VQAs) provide a fruitful toolset to tackle a broad variety of problems, whose solution rely on quantum hardware to estimate quantities that are hard to compute and a classical optimization process to find the optimal parameters for the model. \textbf{b)} The prime factorization problem constitutes one of the corner stones of modern cryptography. Though quantum algorithms are known to factor prime numbers exponentially faster than classically possible, their implementation requires quantum devices of currently infeasible scale and quality. \textbf{c)} We apply a variational method to the factorization problem, tailoring the algorithm to a quantum photonic platform.}
    \label{fig:intro}
\end{figure}
Indeed, quantum technologies promise to address computational problems exponentially faster and more efficiently, when being compared to their classical counterparts, both for research scopes, e.g. simulating complex physical systems \cite{Ayral2023review,McArdle2020review}, as well as for tackling real-world problems \cite{herman2023quantum,fernandez2020crypt}. In this context, two specific quantum computing paradigms that have garnered popularity in recent years are \textit{adiabatic quantum computing} (AQC) and \textit{Variational Quantum Algorithms} (VQAs) \cite{AQCreviewLidar,CerezoVQAreview}. The first draws inspiration from the concept of adiabatic evolution and it maps the problem of interest onto finding the ground state of a given Hamiltonian ($H_f$). Since such a state can be experimentally very complex to achieve, the algorithm starts from the implementation of another state (easier to prepare), which corresponds to the ground state of an initial Hamiltonian $H_i$. Then, the latter is slowly (and in a controlled way) modified, so that the physical system is maintained in the ground state. In the end, when $H_i$ converges to $H_f$, it is possible to retrieve the solution of the considered computational problem.
Despite its robust nature, obtaining fruitful outcomes using AQC on noisy intermediate scale quantum devices (NISQ) is a challenging task, as adiabatically modifying the initial Hamiltonian requires long sequences of quantum gates \cite{messiahQuantum}. 

\begin{figure*}[tb]
    \centering
    \includegraphics[width=0.8\textwidth]{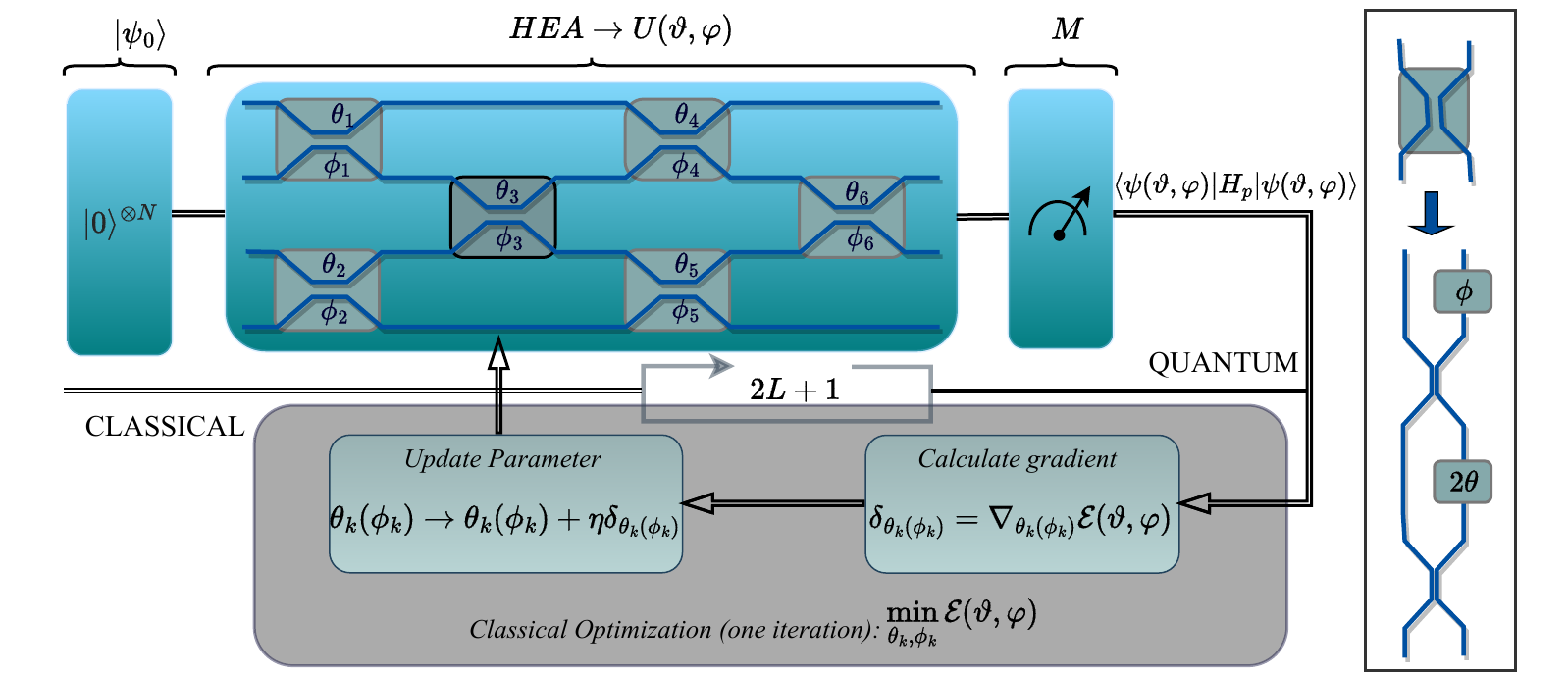}
    \caption{\textbf{Conceptual scheme of the proposed single-photon linear-optics based factorization scheme.} The process is hybrid quantum (above) and classical (below) architecture. The quantum part consists of a tunable photonic circuit (PIC) with four input/output modes \cite{Clements:16}. This device contains  tunable Mach-Zehnder interferometers ($L=6$), with two tunable parameters, $\theta$ and $\phi$, for enabling arbitrary unitary evolutions. The circuit rotates the input state $|\psi_0\rangle$ to $U(\vartheta,\varphi)|\psi_0\rangle$, which is then measured in the computational basis. The classical part covers the calculation of a cost function $\mathcal{E}(\vartheta,\varphi)$ related to the expectation value of the Hamiltonian whose ground state provides the solution to a computational problem (here,  prime factorization of $N=35$). Based on the measurement outcome of $U(\vartheta,\varphi)|\psi_0\rangle$, a classical gradient descent algorithm is used to adapt $U(\vartheta,\varphi)$, minimizing the cost function $\mathcal{E}(\vartheta,\varphi)$. 
The gradients are computed by the forward difference method which involves $2L + 1$ times of function evaluations for each iteration of the optimization process.}
    \label{fig:schematic}
\end{figure*}
Given the limitations of state-of-the-art quantum computers (e.g. low number of qubits and noisy processes that limit circuit depth), a more feasible alternative to AQC algorithms is constituted by VQAs. These rely on classical optimizers to train parametrized quantum circuits to find the solution to computational problems \cite{blekos2023review,TILLY20221}. For instance, such algorithms have been employed in simulating intricate quantum systems to solve quantum chemistry problems \cite{CerezoVQAreview,cao_quantum_2019}, optimizing financial portfolios \cite{brandhofer_benchmarking_2022,Barkoutsos2020improving,Fernández-Lorenzo_2021,Hegade2022portfolio}, and tackling combinatorial optimization challenges \cite{Zhu2022AQAOA,oh2019solving,PhysRevApplied.19.024027}, showcasing their effectiveness in addressing real-world problems.
In this context, one of the most investigated questions was whether this kind of approach could be adopted for the factorization of large numbers, as alternative to Shor's \cite{shor1994algorithms, jiang2018quantum,Peng2008FactorPrl,Hegade2021Fac,karamlou2021analyzing,yan2022factoring}. Indeed, despite being efficient, the implementation of such algorithm is highly  challenging. This question was answered in the affirmative, as it was possible to design an AQC algorithm to tackle this task, which requires a lower number of qubits \cite{peng2008quantum}. In this work, we want to leverage the lower requirements of this algorithm and design a VQA able to solve the same task, while benefitting of the higher experimental feasibility.
From the experimental point of view, VQAs have been implemented on gate-based quantum computers, across platforms such as superconducting and trapped ion quantum computers \cite{Bharati2002review}. However, an alternative to these platforms is constituted by photonics, which allows for remarkable performances when exploiting intrinsic properties of photons. A prominent example is the ultimate demonstration of a quantum advantage \cite{zhong2020quantum, madsen2022quantum} by exploiting the propagation of many photons in complex interferometric networks. This demonstrates that for special-purpose applications that do not require two-qubit gates, which are challenging for photonic systems, such a hardware is very efficient due to its stability, versatility and ability to operate at ambient temperatures. 
VQAs have been already implemented on photonic platforms, tackling specific scientific problems, related to quantum simulations of small molecules or sensing applications, with a very recent implementation tackling a multiparameter case through machine learning \cite{peruzzo2014variational, lee2022error, cimini2024variational}. 
For further developments, a strategy that has been considered aims at tailoring VQA algorithms to photonic architectures via a so-called Hardware-Efficient Ansatz (HEA) \cite{KandalaHEA2017,Benedetti2021,leone2022practical}. This is done by considering the available quantum hardware and optimal resource usage for designing an algorithm, albeit at the expense of some problem-specific expressiveness.

Here, we implement a VQA, which is custom-tailored to photonic platforms, to address a proof-of-principle instance of the factorization problem.  Regarding the adopted platform, we employ four input/output ports of a universal photonic integrated circuit (PIC) consisting of a mesh of programmable interferometers that can implement arbitrary unitary transformations on Fock input states. This kind of platforms is a natural candidate for adaptive computational algorithms, benefitting from their ease of encoding and manipulating information with high fidelity.
Here we present such a protocol where we map our problem to the search of the ground state of a suitable Hamiltonian and implement a gradient based optimization on our photonic platform to obtain the experimental parameters yielding the correct solution. A schematic of this process is depicted in \figref{fig:intro}. Albeit choosing it as a benchmark, our approach's scope goes beyond that task.
Yet, our method for factorization has a better scalability, in terms of the number of required qubits, when being compared to Shor's algorithm and, in the special case of a number whose factors are successive prime numbers, also compared to the adiabatic quantum factoring proposed in \cite{peng2008quantum} (See Supplementary Material Section III). As such, our method is in particular applicable on noisy small and medium-sized quantum hardware, which are within the reach of current technologies. 


\begin{figure}[tb]
    \centering
    \includegraphics[width=0.45\textwidth]{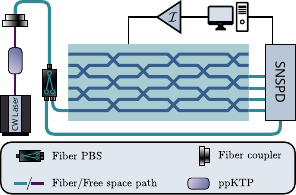}
    \caption{\textbf{Experimental setup.} 
    Orthogonally polarized single-photon pairs generated by a type II Spontaneous Parametric Down-Conversion (SPDC) process are separated using a fiber polarizing beam-splitter (PBS), where one output port is used to herald a pair creation event. The remaining photon is injected into the photonic integrated circuit (PIC) in a fixed spatial mode representing the input state $\ket{\psi_0}=\ket{00}$. The PIC implements a unitary $U(\vartheta, \varphi)$ on the photon-path degree-of-freedom, creating the trial state $\ket{\psi(\vartheta, \varphi)} = U(\vartheta, \varphi) \ket{\psi_0}$. This state is measured in the two-qubit computational basis, where the basis states $\ket{00}$ ($\ket{01}$) are identified by a photon being present in the first (second) output mode, and so on. From the acquired output statistics, the trial states energy $\mathcal{E}(\vartheta, \varphi) = \braket{\psi(\vartheta, \varphi)|H_p|\psi(\vartheta, \varphi)}$ with respect to the Hamiltonian $H_p$ of Eq. \eqref{optHamiltmain} can be computed. A classical gradient descent algorithm is employed to minimize $\mathcal{E}(\vartheta, \varphi)$ by varying the internal and external phases $\vartheta, \varphi$.}
    \label{fig:experiment}
\end{figure}
\section{Results}
\label{sec:res}
The implementation of the factorization problem on the PIC using VQA involves three key steps (a scheme of the approach is depicted in \figref{fig:schematic}). Firstly, through classical preprocessing, we encode the problem into a Ising spin glass Hamiltonian that yields the solution in its ground state. Subsequently, we devise a suitable ansatz by employing a parameterized unitary for the PIC to generate a trial ground state along with a corresponding function to minimize (i.e. loss function). Such a function is designed such that its minimum corresponds to the optimal parameters which solve the task. Finally, we minimize it using a gradient descent optimization to determine the optimal parameters, ultimately providing the ground state. 
For the first step, the prime factor decomposition of a positive integer $N$ into two prime numbers $p$ and $q$, where $p, q \geq 3$ can be mapped into the following optimization problem \cite{Peng2008FactorPrl}:
\begin{equation}
    \underset{x,\,y \,\in \, \mathbb{Z}^+}{\arg\min} \;f(x,y); \;\;\text{where,} \;\; f(x,y) = (N-xy)^2,
    \label{optimal1}
\end{equation}
such that, for optimal values of $x$ and $y$, the function $f(x, y)$ goes to 0, for the values $x_c$ and $y_c$ (see Supplemental Material section II) that correspond to the solution $p$ and $q$ of the factorization problem. 
The values of $N$, $p$, and $q$ are real integers and can be considered as odd numbers without loss of generality. The lengths of the bit-strings corresponding to the binary representation of $x_c$ and $y_c$ are a priori unknown but can be estimated as follows (see Supplemental Material section I):
\begin{equation}
     n_{x,y} = m \left( \mathcal{F} \mp \delta N\right).
     \label{nxnymod1}
\end{equation}
where $\mathcal{F} = \lfloor\sqrt{N}\rfloor$ (with $\lfloor a \rfloor$ indicating the largest integer smaller or equal to $a$) and $m(b)$ denotes the smallest number of bits required for representing $b$. The offset $\delta N$, instead, is required because the gap between two successive primer numbers is not constant and does not grow linearly with $N$ \cite{hardy1979introduction} and it amounts to $\sqrt{\mathcal{F}^2-N}$ (see Supplemental Material section I). 
This implies that the solution can be expressed in the following as a binary number with the maximal bit-string length upper bounded by $n_x$, for the first factor, and $n_y$ for the second: 
\begin{equation}
x_c = x_{n_x - 1}\dots x_1 x_0, \;\;\;\;\;\;\;\;  y_c = y_{n_y} - 1\dots y_1 y_0.
\label{bitstrings}
\end{equation}
However, when the number $N$ to factorize is given by the product of two prime numbers, the number of unknown bits is lower, if we consider that each odd binary number ends with $1$. Therefore, $n_x$ ($n_y$) is reduced to $n_x-1$ ($n_y-1$) and, if we want to encode $x_c$ and $y_c$ in a quantum state, the required Hilbert space dimension results $2^{n_x + n_y - 2}$. Also in this case, these bit-lengths are an overestimation of the real length of the factors.

In our implementation, anyway, for practical purposes we can reduce the number of unknown bits even further. Indeed, if we consider the factorization of the number $35=5 \times 7$, which is equal to the largest number factorized through a simplified version of Shor's algorithm \cite{amico2019experimental}, which exploited 7 qubits and succeeded roughly $14\%$ of the cases, the length of the bit-strings $x$ and $y$ constituting the prime factors must amount to $n_x=n_y=3$ (as detailed in \cite{Peng2008FactorPrl} and as reported in Supplemental Material section I). This implies that also the most significant bit of the factors must be 1, otherwise a lower number of bits would be sufficient. So, in this specific case, the total number of unknown bits is further reduced, because, our factors $x_c$ and $y_c$ will have the form $1x_11$ and $1y_11$ respectively. Let us note that this restriction holds only in our case, but the protocol can handle more general cases. Indeed, it is possible to evaluate the maximum number of required bits (as reported in Supplemental Material section I), which will be the number of bits outputted by the method. Hence, if a lower amount of bits is sufficient, the most significant ones will be 0. 
In the present scenario, the Hamiltonian that encodes the problem can be written in an Ising-like form as follows: 
\begin{equation}
    H_P = \left[NI- \left(4 I + 2 \Pi_z^1 + I)\right)_1\otimes\left(4 I + 2\Pi_z^1 + I\right)_2\right]^2,
    \label{optHamiltmain}
\end{equation}  
 where $N$ is the number to factorize, $I$ is the identity operator, $\Pi_z^1$  refers to the projector onto the eigenvector of the $\sigma_Z$ operator corresponding to outcome $1$ and the subscript indicates the qubit on which it acts (see the Supplemental Material section II for further details). Then, the state that will minimize Eq.~\eqref{optHamiltmain} will be represented by a two-qubit state $\ket{x_1 y_1}$, where the two unknown bits will be $x_1$ and $y_1$. This is due to the form of the Hamiltonian, because the two factors $x$, $y$ are expressed, in binary form, as follows: $4+2x_1+1$ and $4+2y_1+1$. So, the ground state of $H_p$ will be the one for which the expected values of the projector onto the eigenstate $|1\rangle$ of the operator $\sigma_z$ amount to $x_1$ and $y_1$. 

At this point, we need to design a suitable ansatz, so we consider a fixed two-qubit initial state $\ket{\psi_0}$ which is rotated by some parameterized unitary $U(\vartheta, \varphi)$ to a trial state $\ket{\psi(\vartheta, \varphi)} = U (\vartheta, \varphi) \ket{\psi_0}$. This state is then projected on the Z basis and the achieved statistics is employed for computing the expectation value of the energy $\mathcal{E}(\vartheta,\varphi)$, as follows:
\begin{equation}
\label{eq:energy_unitary1}
    \mathcal{E}(\vartheta,\varphi) = \bra{\psi_0}U^{\dag} (\vartheta, \varphi) H_p U (\vartheta, \varphi) \ket{\psi_0}.
\end{equation}
According to the variational principle, such an energy value must be greater than or equal to the ground state energy $E_0$. Hence, $\mathcal{E}(\vartheta, \varphi)$ constitutes the cost function to be minimized, by adaptively changing the parameters of the implemented unitary.
Given these premises, the solution boils down to finding the state that corresponds to the ground state of the Hamiltonian $H_p$. This implies that our goal is to find the unitary $U$ such that $U(\vartheta, \varphi)|\psi_0\rangle=|x_1,y_1\rangle$. Due to the commutativity property of multiplication, it follows that $N=p\times q=q\times p$. This implies that there are two unitaries encoding the correct solution of our problem, as the solution can be given by $U(\vartheta_1, \varphi_1)|\psi_0\rangle=|01\rangle$ and $U(\vartheta_2, \varphi_2)|\psi_0\rangle=|10\rangle$. 
Hence, for the solution $|01\rangle$, the two factors will be $x_c=101=5$ and $y_c=111=7$, while for $|10\rangle$, they will amount to $x_c=111=7$ and $y_c=101=5$, see \figref{fig:schematic}.

From an experimental point of view, we prepare the state $\ket{\psi_0}$ by harnessing the path degree of freedom in four spatial modes of a PIC, following the encoding
$|\psi_0\rangle= \ket{1,0,0,0} = \ket{00}, \ket{\psi_1} = \ket{0,1,0,0}= \ket{01}, \ket{\psi_2}= \ket{0,0,1,0} = \ket{10}$ and $\ket{\psi_3}= \ket{0,0,0,1}= \ket{11}$, 
where $\ket{1,0,0,0}$ denotes a photon populating the first mode of the PIC, and so on.
\begin{figure}
    \centering    \includegraphics[width=0.45\textwidth]{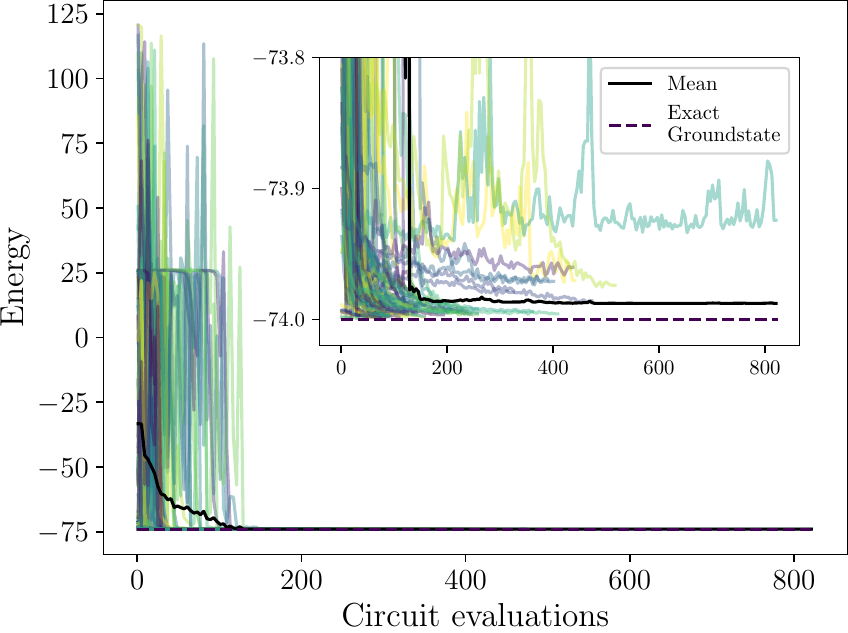}
    \caption{\textbf{Convergence of the estimated ground state energy for different initial configurations}. Calculated energy $\braket{\Psi | H_p | \Psi}$ (Eq. \eqref{eq:energy_unitary1}) at each optimization step, for $117$ different initial configurations of the experiment (colored lines), as they approach the theoretically predicted ground state energy $\mathrm{E_g}$ (dashed line). The configuration at each step consists in the set of internal and external phases $(\vartheta, \varphi)$ applied to the integrated circuit, which amounts to a unitary operation $U(\vartheta, \varphi)$. Accordingly, at the $i$th step, the output state will be $|\Psi_i\rangle=U(\vartheta_i, \varphi_i)|\psi_0\rangle$, where $|\psi_0\rangle$ is the initial input. To evaluate the obtained energy, due to the form of our Hamiltonian $H_p$ (see Eq. \eqref{optHamiltmain}), it is sufficient to post-process the statistics obtained by projecting the output state onto the computational basis. The black line shows the mean experimental energy at every step, where experiments with a final energy $\mathcal{E} \geq 0$ have been removed (that is excluding \num{5} out of \num{117} repetitions). A magnified region close to $\mathrm{E_g}$ is depicted in the inset. The theoretically predicted ground state energy is reached by the optimization procedure after 27, 28, 115 steps within a relative uncertainty of 1\%, 0.1\%, and 0.01\% respectively. Let us note that the achieved energy values are lower than 0, because we are removing a constant term from the Eq. \eqref{optHamiltmain}.}
    \label{fig:results_cost_per_run}
\end{figure}
Such a single photon input is generated by type-II Spontaneous parametric down conversion (SPDC) in a periodically poled $\mathrm{KTiOPO_4}$-crystal, with a poling period of \SI{46.2}{\micro\meter}, in a single-pass configuration as depicted in \figref{fig:experiment}, generating wavelength degenerate photons at $\lambda_s=\lambda_i=\SI{1550}{\nano\meter}$ out of a $\SI{775}{\nano\meter}$ continuous wave pump. 
Then, one photon is directly sent to a detection unit consisting of Superconducting Nanowire Single Photon Detectors (SNSPDs) with a detection efficiency of \SI{\sim 95}{\percent} and a dark-count rate of \SI{\sim 300}{\hertz}.
It thereby heralds the presence of its twin, which is routed to a tunable optical circuit. This is the core of our experimental apparatus and it is
constituted by a six-mode universal PIC, featuring only linear optical elements, i.e. phase shifters and beam splitters, whose architecture follows the optimal design proposed in \cite{Clements:16}.
In detail, the circuit consists of a grid of variable beam splitters, i.e. Mach-Zehnder interferometers (see \figref{fig:schematic}), each characterized by a transformation matrix denoted as $T_{jk}(\theta, \phi)$, where $\theta$ determines the reflectivity and $\phi$ is an external phase. With a specific ordered sequence $S$ of these transformation matrices, it is possible to create a programmable unitary operator described as: 
\begin{equation}
    U (\vartheta, \varphi) = D \prod_{j,k \in S, l} T_{jk} (\theta_l, \phi_l); \quad l = \{1,2,\dots L\},
    \label{HEA}
\end{equation}
where $D$ is some arbitrary diagonal matrix. It is worth noting that the sets of internal phases $\vartheta = \{ \theta_1, \theta_2,....\theta_L \}$ and external ones $\varphi = \{ \phi_1, \phi_2,....\phi_L \}$ 
amounts to $2L$ parameters, where $L$ is the number of Mach-Zehnder interferometers.
This structure allows us to perform arbitrary unitary operations on six-dimensional input Fock states, by properly choosing the values of the phase shifts. 


The PIC is manufactured in a borosilicate glass substrate through femtosecond laser waveguide writing \cite{corrielli2021femtosecond}. The tunability is achieved by fabricating thermo-optic phase shifters (i.e. resistive heaters) in a thin film of gold deposited on the chip surface. The device exhibits an overall transmission of $54\%$ and can implement arbitrary unitary operations on six modes with an average amplitude fidelity of 0.9970 ± 0.0017 \cite{pentangelo2024high}. While the employed PIC has a negligible polarization extinction ratio (PER), the path encoding scheme solely relies on a homogenious PER along the PICs width to maintain interference.

For our experiment, 
we need to use only 
four input and output modes, that are tunable by acting on twelve phase shifters (L=6), while the surrounding actuators are set in reflection ($\theta = \pi/2$) to isolate these modes from the others available in the PIC. Our ansatz is custom-tailored to these structures.
Hence, to find these two sets of optimal parameters $(\vartheta_1, \varphi_1)$ and $(\vartheta_2, \varphi_2)$, we perform $117$ repetitions of the experiment, starting with random initial parameters. 
For each repetition, we initialize our ansatz by applying twelve random phase shifts ($L=6$) and injecting the input state as $|\psi_0\rangle=|00\rangle$, namely a single photon in the first mode. Then, we collect the photon statistics at each output, corresponding to a projection of the rotated input onto the computational basis, comprising the states $|00\rangle$, $|01\rangle$, $|10\rangle$ and $|11\rangle$. Following this, we post-process the data to evaluate the energy 
of $\ket{\psi_0}$ with respect to $H_P$ on a classical computer, using Eq.~\eqref{eq:energy_unitary1}. 
Note that, for $H_p$, which only involves expected values in the $\sigma_z$ basis, a full tomography on the output state is not required to achieve the correct value of Eq.~\eqref{eq:energy_unitary1} (see the Supplemental Material section IV).

For each of the 117 repetitions of the experiment, we carry out an optimization, through a gradient descent algorithm, made of several iterations. For each step of the optimization, we derive a new set of phases, which is subsequently applied to the circuit through a classical feedback loop. The gradients for each of the $2L$ parameters is obtained using a forward finite difference method. This requires a total number 2L+1 of circuit evaluations to be performed in each feedback, to update all the phases. Every iteration $i$ yields a new energy value $\mathcal{E}(\vartheta_i,\varphi_i)$ and the process stops when it converges to a solution. This implies that the variation between the energies of two consecutive steps falls below a specified threshold $\epsilon = 10^{-4}$. At this point, the obtained photon statistics of the last iteration constitutes an approximation of the ground state energy and therefore represents the solution. 

On average, one optimization step is performed every \SI{174}{\second}, of which \SI{42}{\second} are spent acquiring detection events. The remaining time can be attributed to the operation of the PIC. It is worthwhile to point out that the classical optimization algorithm, in principle, requires no knowledge about $U(\vartheta, \varphi)$ or even the possibly nonlinear mapping between the phases and their control currents. In our case, though, we exploited a pre-characterization of the device, to ensure that no light would go to the undetected modes of the 6 input/ 6 output circuit. 

\begin{figure*}
    \centering
    \includegraphics[width=0.95\textwidth]{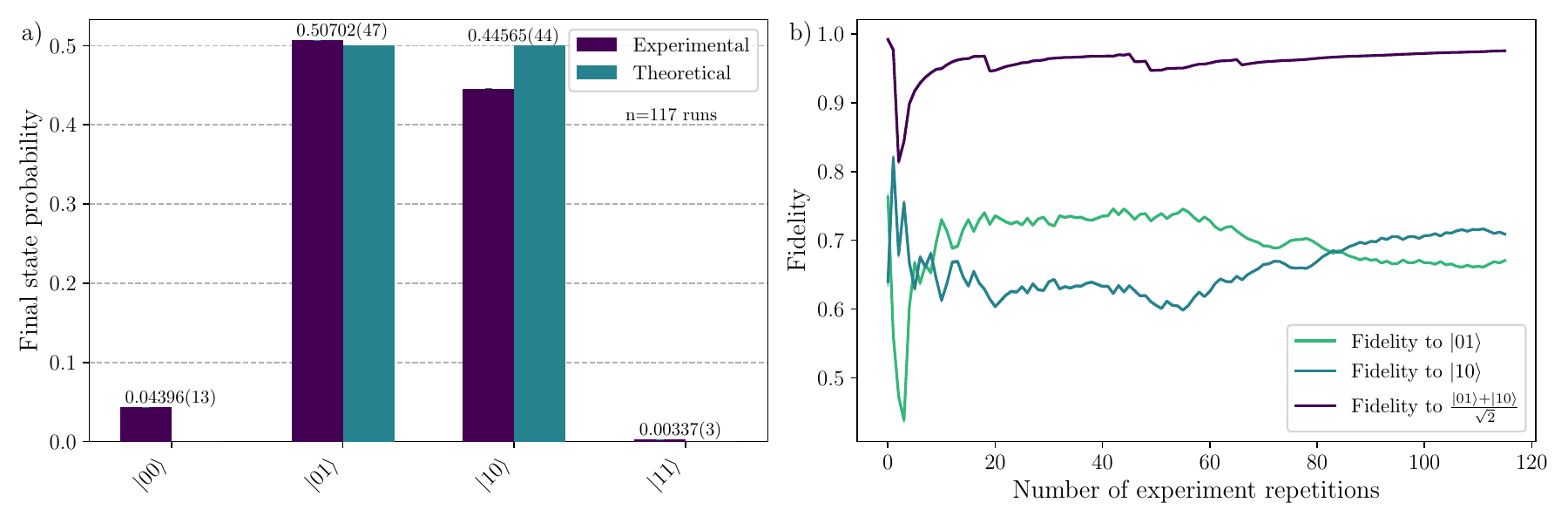}
    \caption{\textbf{Experimental factorization results.} \textbf{a)} The bars correspond to the experimental frequencies with which a photon is detected at each of the four outputs of our circuit. The reported values are obtained by averaging over the results of \num{117} different experimental optimizations, starting from different configurations. For our encoding, injecting one photon in the $i$-th mode constitutes the $i$-th element of a 2-qubits computational basis. Therefore, the bars representing the highest frequencies indicate the ground state $\ket{x_1 y_1}$ of the Hamiltonian outlined in Eq.~\eqref{optHamiltmain}, which is an almost equal superposition of $\ket{01}$ and $\ket{10}$. This effectively resolves the factorization problem, allowing us to construct the factors as $1 x_1 1 = 101 =5 $ and $1 y_1 1 = 111 =7 $, or vice versa. 
    \textbf{b)} As all states of the form $\sqrt{\alpha}\ket{01} + \sqrt{1-\alpha}\ket{10}$ exhibit the same energy with respect to $H_p$ (see Supplemental Material section IV), every individual optimization converges to one arbitrary of those possible outcomes. The average output of independent repetitions thus approaches the equal superposition $\alpha=0.5$, as can be seen on the fidelity of $0.97546(5)$ towards this state after 117 repetitions (violet), while the fidelities to both $\ket{01}$ (green) and $\ket{10}$ (teal) concordantly revolve around $\num{0.7}$. The fidelity (Eq. \eqref{equ:fidelity}) is evaluated by interpreting the output distribution measured in the computational basis as normalized state vector. For both plots, the uncertainties (shaded regions for 2 standard deviations), evaluated considering an underlying Poissonian statistics, using a Monte Carlo Simulation and are too small to be visible.}
    \label{fig:results_quasiprobability}
\end{figure*}

The convergence of the energy value with respect to the function evaluations is depicted in \figref{fig:results_cost_per_run}, where the different curves correspond to different initial phase configurations and the dashed line represents the theoretical limit. It can be clearly seen that the optimization works smoothly and is able to reach the ground state in every instance with an accuracy ranging from $99\%$ to $99.99\%$. Furthermore, let us point out that the ground state of $H_p$ is two-fold degenerate and each instance of the experiment converges to one particular ground state corresponding to one optimal unitary i.e., either $U(\vartheta_1, \varphi_1)$ or $U(\vartheta_2, \varphi_2)$ .
This is visible throughout the 117 repetitions of the experiment, whose averaged probabilities are represented in \figref{fig:results_quasiprobability}a (purple bars), where we also compare them to the theoretical predictions (in blue). We obtained the solutions $\ket{01}$ and $\ket{10}$ with $50.7\%$ and $44.6\%$ respectively. This implies, as previously mentioned, that the unknown bits for our two factors are $ \ket{x_1y_1}= \ket{10}$ or, equivalently, $\ket{01}$.

This is also reconfirmed in 
\figref{fig:results_quasiprobability}b, where we plot the 
fidelities
\begin{equation}
    F(\rho, \sigma) = \tr \left( \sqrt{\sqrt{\rho}\sigma\sqrt{\rho}} \right),
    \label{equ:fidelity}
\end{equation}
between the state $\sigma$, which amounts to $\sqrt{|\alpha|} |0\rangle + \sqrt{(1-|\alpha|)} |1\rangle$, at each experimental repetition for the last optimization step, and the target state $\rho$. Note that the form of state $\sigma$ is obtained by measuring in the Pauli $\sigma_Z$  basis. Across the \num{117} experimental repetitions, the measured state emerges as $\frac{1}{\sqrt{2}}(\ket{01}+\ket{10})$ with a final fidelity close to unity (depicted in purple). Accordingly, the fidelities to the individual solution states $\ket{01}, \ket{10}$ are close to $1/\sqrt{2}$ (depicted in blue, green).


\section{Discussion}
\label{sec:conclusion}
This work reports on the implementation of a VQA on an integrated photonic circuit, exploiting  an ansatz which is tailored to our platform. In particular, we present a proof-of-principle demonstration for the factorization of $35 = 5 \times 7$, executed on a four-input/output modes of a universal PIC. Our strategy involves the use of a qudit with four levels, corresponding to the four modes of the photonic circuit. This is then interpreted as two qubits, encoding the solution in the ground state of an Ising Hamiltonian. Subsequently, we determine its energy through a hybrid quantum-classical optimization problem, utilizing gradient descent. In further detail, we design an ansatz specifically for state-of-the-art PICs that allow high-quality processing of single photons by using only passive optical elements 
\cite{Clements:16}. Our encoding of the two-qubit computational basis involves the binary representation of the mode from which one single photon enters an optical circuit, where the first mode corresponds to $|00\rangle$, the second to $|01\rangle$, and so forth.
At each step of the optimization, we collect the statistics obtained by projecting the evolved input state onto the computational basis and post-process it to retrieve the energy value. Then, we identify a new configuration of our circuit and we iterate the procedure, until it converges to the minimum energy value, corresponding to the ground state of our Hamiltonian. This allows us to reach the correct solution for our factorization problem with a success probability exceeding 98\%, in fewer than 150 iterations and for $117$ different starting conditions, highlighting its reduced sensitivity to initial parameters.
 This study marks the first instance of factorization on an optical computing device and shows the utility of the photonic devices in Variational Quantum Algorithms, showcasing advantages in terms of resource requirements and resilience against inherent noises. This is reflected by the fact that, with the integrated platforms currently available on the market, our implementation could already factorize a semi-prime number up to $93$, while considering the product of two consecutive primes, this limit goes up to $899$. These numbers are well beyond what has current been achieved with Shor's algorithm \cite{amico2019experimental} and this limit is likely to increase with recent developments in integrated photonics. Indeed, until now, the largest number factorized through a simplified version of Shor's algorithm is $35$, by exploiting 7 qubits and obtained a success rate of roughly $14\%$ of the cases \cite{amico2019experimental}. The number of required qubits was then lowered down in a more recent work, where $21$ was factored using 5 qubits \cite{skosana2021demonstration}. Moreover, although our demonstration is dedicated to the factorization problem, we want to emphasize that our approach can readily be employed to address other challenges, given that our results indicate that universal integrated circuits are a suitable platform for the implementation of VQAs. 
 Let us note that the integrated photonic platforms that are currently on the market are already sufficient to encode up to 5 qubits using path-encoded single-photon architectures. This places our hardware-efficient variational framework well within the range of many established benchmark problems in near-term quantum computing. For instance, ground-state estimation of simple molecular systems such as $\mathrm{H_2}$ using VQE requires only few qubits \cite{kandala2017hardware, o2016scalable}, and combinatorial problems like MaxCut on 3 to 4 node graphs can also be encoded on systems smaller than 5 qubits \cite{farhi2014quantum}. These examples are representative of diverse problem classes and underscore that our method is not inherently limited to factorization tasks, but is broadly applicable within the capabilities of current photonic hardware. Moreover, the versatility, high achievable fidelity in the implementation of arbitrary unitary transformations and scalability of photonics platforms make them particularly apt for adaptive protocols, e.g. the training of machine learning algorithms. On the other hand, the limitations of this platform are related to the difficulty of implementing two-qubit gates and hence non-linear operations. Indeed, the fact that photons tend not to interact with the environment implies that two-qubit gates can only be implemented probabilistically and heralded through ancillary photons or post-selection \cite{knill2001scheme}. 
These issues were circumvented in our case by encoding $n$ qubits in $2^n$ modes, offering a platform which is exactly equivalent to the quantum circuit formalism. While this approach offers the advantage of the number of input photons staying constant - specifically one - independent of problem size, it exhibits the drawback of the number of necessary PIC modes scaling exponentially in the length of the encoded bit strings. Consequently, the number of parameters to optimize also increases exponentially. Furthermore, the reliance on a single photon relaxes requirements on PIC transmittivity and detection efficiency. As these quantities affect the integration time exponentially in the number of photons, they only impact our protocol in a constant manner. 
However, the drawbacks can potentially be eased by future developments, as suggested by several recent studies investigating the possibility of implementing non-linear operations through encoding, feedback loops and enhancing the expressivity of the circuit through quantum interference \cite{innocenti2023potential, spagnolo2022experimental, govia2022nonlinear}. Moreover, recent studies have highlighted that optical computational models could grant lower energy consumption, with respect to standard models \cite{hamerly2019large}.
\section{Methods}
\label{sec:method}
{\bf Hardware-Efficient Ansatz tailored to photonic platforms.}
Recently, there have been several studies, particularly focused on realizing various computing tasks like deep learning and generative networks on the PICs 
\cite{Mengu2022deeplearning,Wang23photonicLearning}. However, our approach primarily resorts to the multiport interferometric design of the PICs introduced in \cite{Clements:16}. As previously outlined, our implementation of the factorization algorithm involves the use of four spatial modes (see~\figref{fig:schematic}). 
In general, for an interferometer with $M$ modes, full connectivity and expressibility is achievable through a total of $M(M-1)/2$ variable beam splitters,(in our case $6$). Each variable beam splitter is constructed using a Mach-Zehnder Interferometer comprising two $50:50$ directional couplers with tunable internal and external phases. The action of the circuit can be represented by a sequence of $4\times 4$ unitary matrices denoted as $T_{j,k}(\theta,\phi)$, with nonzero elements in the $j$th and $k$th row and column representing $j$th and $k$th mode with $0\leq\theta\leq \pi/2$ and $0\leq\phi\leq 2\pi$. Our ansatz can be broken down into such a sequence $S$ of the $T_{j,k}(\theta,\phi)$ matrices, which, for a given set of $(\vartheta, \varphi)$, can diagonalize any provided unitary matrix $U$. In our structure, depicted in the quantum segment of~\figref{fig:schematic}, we employ the following sequence:
\begin{align}\nonumber
    D = T_{3,4}(\theta_6,\phi_6) &T_{2,3}(\theta_5,\phi_5) U T^{-1}_{1,2}(\theta_1,\phi_1)\dots\\
    &T^{-1}_{3,4}(\theta_2,\phi_2)T^{-1}_{2,3}(\theta_3,\phi_3)T^{-1}_{1,2}(\theta_4,\phi_4),
\end{align}
where $D$ is a diagonal matrix. It is crucial to emphasize that a specific order in the phases is essential to achieve the necessary expressibility in the ansatz. The subsequent derivation yields a general parameterized unitary that can be implemented in the interferometer as,
\begin{align}\nonumber
     U = \Tilde{D}T_{2,3}(\theta_5,\phi_5)&T_{3,4}(\theta_6,\phi_6)T_{1,2}(\theta_4,\phi_4) \dots\\
     &T_{2,3}(\theta_3,\phi_3)T_{3,4}(\theta_2,\phi_2) T_{1,2}(\theta_1,\phi_1),
\end{align}
as presented in a compact form in Eq.~\eqref{HEA}. $\Tilde{D}$ represents a single-mode phase matrix associated with a specific set of $(\vartheta, \varphi)$ and follows the relation $\Tilde{D} T_{j,k} = T^{-1}_{j,k} D$. However, in our experimental setup, where $U = U(\vartheta, \varphi)$ is a parameterized unitary to be optimized subsequently using a classical optimizer, we can conveniently choose $\Tilde{D}$ as an Identity matrix without any loss of generality. 

\textbf{Classical optimization.}
In our protocol, we apply $U(\vartheta, \varphi)$ to a single photon initial state and measure the resulting trial state $U(\vartheta, \varphi)|\psi_0\rangle$ at the output of the PIC. Subsequently, we compute a parameterized cost function $\mathcal{E}(\vartheta, \varphi)$, as outlined in Eq.~\eqref{eq:energy_unitary1}, based on the gathered photon statistics. This cost function is then subjected to classical optimization, as depicted in the classical segment of~\figref{fig:schematic}.
The approach discussed in this study differs slightly from the commonly used variational methods that rely on conventional gate models. Achieving a satisfactory solution with HEA often requires the utilization of quantum circuits with substantial depth ans intricate parameterization, involving an increased number of layers. Nevertheless, within the depicted PIC, the qumodes are fully interconnected, which allows for expanding the entire Hilbert space, particularly with the encoding method selected for implementing the ansatz. Consequently, Clement's structure ensures full expressiveness in order to find the solution by classical optimization. 
The most prevalent method for optimizing an objective function involves computing its gradients, representing the change in the function concerning variations in its parameters. This approach is adopted in our case to minimize $\mathcal{E}(\vartheta, \varphi)$. Here, we compute the gradients using a finite forward difference scheme, which is expressed as follows:
\begin{equation}
    \nabla_{\theta_k (\phi_k)} \mathcal{E}(\vartheta,\varphi)   =   \frac{1}{h}\bigg( \mathcal{E}(\theta_k (\phi_k)+h) - \mathcal{E}(\vartheta,\varphi) \bigg),
\end{equation}
and the corresponding parameters are updated by $\theta_k (\phi_k) \rightarrow \theta_k (\phi_k) + \eta \nabla_{\theta_k (\phi_k)} \mathcal{E}(\vartheta,\varphi)$ where $\eta$ is the learning rate. The spacing $h$ is chosen as $\num{0.01}$, and the learning rate $\eta$ to be $\num{0.03}$. It is crucial to emphasize that, in every iteration of the optimization process, the gradients for each parameter must be calculated individually. Therefore, each gradient calculation for a specific parameter necessitates a single measurement, resulting in a total of $(2L+1) \times n_{it}$ number of measurements to obtain the final optimal parameters with $n_{it}$ being the number of iterations.

\section*{Acknowledgements}
\label{sec:Acknowledgement}
This research was co-funded by the European Union (HORIZON Europe Research and Innovation Programme, EPIQUE, No 101135288). Views and opinions expressed are however those of the author(s) only and do not necessarily reflect those of the European Union or the European Commission-EU. Neither the European Union nor the granting authority can be held responsible for them. This project has received funding from the European Union’s Horizon 2020 research and innovation programme under grant agreement No 899368 (EPIQUS), under the Marie Skłodowska-Curie grant agreement No 956071 (AppQInfo) and under Grant Agreement no. 101017733 (QuantERA II).
This research was funded in whole or in part by the Austrian Science Fund (FWF)[10.55776/ESP205] (PREQUrSOR), [10.55776/I6002] and [10.55776/F71] (BeyondC). For open access purposes, the author has applied a CC BY public copyright license to any author accepted manuscript version arising from this submission. The financial support by the Austrian Federal Ministry of Labour and Economy, the National Foundation for Research, Technology and Development and the Christian Doppler Research Association is gratefully acknowledged.
 K.P. and X.C. also appreciate the funding from the Basque Government through Grant No. IT1470-22, the project grant PID2021-126273NB-I00 funded by MCIN/AEI/10.13039/501100011033 and by ``ERDF A way of making Europe" and ``ERDF Invest in your Future", and the Spanish Ministry of Economic Affairs and Digital Transformation through the QUANTUM ENIA project call-Quantum Spain project.

\section*{Author contributions}
$ ^*$ These authors have contributed equally. P.S and I.A. designed and conducted the experiment. K.P. developed the theory, discussing with Y.B.. S.S. built the single photon source. S.S. and Z.Y. participated in the experiment. C.P., S.P., A.C., F.C. and R.O. conducted the design, fabrication and calibration of the integrated photonic processor. X.C. and P.W. supervised the project. All authors discussed the results and reviewed the manuscript.

\section*{Competing interest}
The authors declare no competing interests.


\begin{thebibliography}{49}%
\makeatletter
\providecommand \@ifxundefined [1]{%
 \@ifx{#1\undefined}
}%
\providecommand \@ifnum [1]{%
 \ifnum #1\expandafter \@firstoftwo
 \else \expandafter \@secondoftwo
 \fi
}%
\providecommand \@ifx [1]{%
 \ifx #1\expandafter \@firstoftwo
 \else \expandafter \@secondoftwo
 \fi
}%
\providecommand \natexlab [1]{#1}%
\providecommand \enquote  [1]{``#1''}%
\providecommand \bibnamefont  [1]{#1}%
\providecommand \bibfnamefont [1]{#1}%
\providecommand \citenamefont [1]{#1}%
\providecommand \href@noop [0]{\@secondoftwo}%
\providecommand \href [0]{\begingroup \@sanitize@url \@href}%
\providecommand \@href[1]{\@@startlink{#1}\@@href}%
\providecommand \@@href[1]{\endgroup#1\@@endlink}%
\providecommand \@sanitize@url [0]{\catcode `\\12\catcode `\$12\catcode `\&12\catcode `\#12\catcode `\^12\catcode `\_12\catcode `\%12\relax}%
\providecommand \@@startlink[1]{}%
\providecommand \@@endlink[0]{}%
\providecommand \url  [0]{\begingroup\@sanitize@url \@url }%
\providecommand \@url [1]{\endgroup\@href {#1}{\urlprefix }}%
\providecommand \urlprefix  [0]{URL }%
\providecommand \Eprint [0]{\href }%
\providecommand \doibase [0]{https://doi.org/}%
\providecommand \selectlanguage [0]{\@gobble}%
\providecommand \bibinfo  [0]{\@secondoftwo}%
\providecommand \bibfield  [0]{\@secondoftwo}%
\providecommand \translation [1]{[#1]}%
\providecommand \BibitemOpen [0]{}%
\providecommand \bibitemStop [0]{}%
\providecommand \bibitemNoStop [0]{.\EOS\space}%
\providecommand \EOS [0]{\spacefactor3000\relax}%
\providecommand \BibitemShut  [1]{\csname bibitem#1\endcsname}%
\let\auto@bib@innerbib\@empty
\bibitem [{\citenamefont {Ayral}\ \emph {et~al.}(2023)\citenamefont {Ayral}, \citenamefont {Besserve}, \citenamefont {Lacroix},\ and\ \citenamefont {Ruiz~Guzman}}]{Ayral2023review}%
  \BibitemOpen
  \bibfield  {author} {\bibinfo {author} {\bibfnamefont {T.}~\bibnamefont {Ayral}}, \bibinfo {author} {\bibfnamefont {P.}~\bibnamefont {Besserve}}, \bibinfo {author} {\bibfnamefont {D.}~\bibnamefont {Lacroix}},\ and\ \bibinfo {author} {\bibfnamefont {E.~A.}\ \bibnamefont {Ruiz~Guzman}},\ }\href {https://doi.org/10.1140/epja/s10050-023-01141-1} {\bibfield  {journal} {\bibinfo  {journal} {The European Physical Journal A}\ }\textbf {\bibinfo {volume} {59}},\ \bibinfo {pages} {227} (\bibinfo {year} {2023})}\BibitemShut {NoStop}%
\bibitem [{\citenamefont {McArdle}\ \emph {et~al.}(2020)\citenamefont {McArdle}, \citenamefont {Endo}, \citenamefont {Aspuru-Guzik}, \citenamefont {Benjamin},\ and\ \citenamefont {Yuan}}]{McArdle2020review}%
  \BibitemOpen
  \bibfield  {author} {\bibinfo {author} {\bibfnamefont {S.}~\bibnamefont {McArdle}}, \bibinfo {author} {\bibfnamefont {S.}~\bibnamefont {Endo}}, \bibinfo {author} {\bibfnamefont {A.}~\bibnamefont {Aspuru-Guzik}}, \bibinfo {author} {\bibfnamefont {S.~C.}\ \bibnamefont {Benjamin}},\ and\ \bibinfo {author} {\bibfnamefont {X.}~\bibnamefont {Yuan}},\ }\href {https://doi.org/10.1103/RevModPhys.92.015003} {\bibfield  {journal} {\bibinfo  {journal} {Rev. Mod. Phys.}\ }\textbf {\bibinfo {volume} {92}},\ \bibinfo {pages} {015003} (\bibinfo {year} {2020})}\BibitemShut {NoStop}%
\bibitem [{\citenamefont {Herman}\ \emph {et~al.}(2023)\citenamefont {Herman}, \citenamefont {Googin}, \citenamefont {Liu}, \citenamefont {Sun}, \citenamefont {Galda}, \citenamefont {Safro}, \citenamefont {Pistoia},\ and\ \citenamefont {Alexeev}}]{herman2023quantum}%
  \BibitemOpen
  \bibfield  {author} {\bibinfo {author} {\bibfnamefont {D.}~\bibnamefont {Herman}}, \bibinfo {author} {\bibfnamefont {C.}~\bibnamefont {Googin}}, \bibinfo {author} {\bibfnamefont {X.}~\bibnamefont {Liu}}, \bibinfo {author} {\bibfnamefont {Y.}~\bibnamefont {Sun}}, \bibinfo {author} {\bibfnamefont {A.}~\bibnamefont {Galda}}, \bibinfo {author} {\bibfnamefont {I.}~\bibnamefont {Safro}}, \bibinfo {author} {\bibfnamefont {M.}~\bibnamefont {Pistoia}},\ and\ \bibinfo {author} {\bibfnamefont {Y.}~\bibnamefont {Alexeev}},\ }\href {https://doi.org/10.1038/s42254-023-00603-1} {\bibfield  {journal} {\bibinfo  {journal} {Nature Reviews Physics}\ }\textbf {\bibinfo {volume} {5}},\ \bibinfo {pages} {450} (\bibinfo {year} {2023})}\BibitemShut {NoStop}%
\bibitem [{\citenamefont {Fernández-Caramès}\ and\ \citenamefont {Fraga-Lamas}(2020)}]{fernandez2020crypt}%
  \BibitemOpen
  \bibfield  {author} {\bibinfo {author} {\bibfnamefont {T.~M.}\ \bibnamefont {Fernández-Caramès}}\ and\ \bibinfo {author} {\bibfnamefont {P.}~\bibnamefont {Fraga-Lamas}},\ }\href {https://doi.org/10.1109/ACCESS.2020.2968985} {\bibfield  {journal} {\bibinfo  {journal} {IEEE Access}\ }\textbf {\bibinfo {volume} {8}},\ \bibinfo {pages} {21091} (\bibinfo {year} {2020})}\BibitemShut {NoStop}%
\bibitem [{\citenamefont {Albash}\ and\ \citenamefont {Lidar}(2018)}]{AQCreviewLidar}%
  \BibitemOpen
  \bibfield  {author} {\bibinfo {author} {\bibfnamefont {T.}~\bibnamefont {Albash}}\ and\ \bibinfo {author} {\bibfnamefont {D.~A.}\ \bibnamefont {Lidar}},\ }\href {https://doi.org/10.1103/RevModPhys.90.015002} {\bibfield  {journal} {\bibinfo  {journal} {Rev. Mod. Phys.}\ }\textbf {\bibinfo {volume} {90}},\ \bibinfo {pages} {015002} (\bibinfo {year} {2018})}\BibitemShut {NoStop}%
\bibitem [{\citenamefont {Cerezo}\ \emph {et~al.}(2021)\citenamefont {Cerezo}, \citenamefont {Arrasmith}, \citenamefont {Babbush}, \citenamefont {Benjamin}, \citenamefont {Endo}, \citenamefont {Fujii}, \citenamefont {McClean}, \citenamefont {Mitarai}, \citenamefont {Yuan}, \citenamefont {Cincio},\ and\ \citenamefont {Coles}}]{CerezoVQAreview}%
  \BibitemOpen
  \bibfield  {author} {\bibinfo {author} {\bibfnamefont {M.}~\bibnamefont {Cerezo}}, \bibinfo {author} {\bibfnamefont {A.}~\bibnamefont {Arrasmith}}, \bibinfo {author} {\bibfnamefont {R.}~\bibnamefont {Babbush}}, \bibinfo {author} {\bibfnamefont {S.~C.}\ \bibnamefont {Benjamin}}, \bibinfo {author} {\bibfnamefont {S.}~\bibnamefont {Endo}}, \bibinfo {author} {\bibfnamefont {K.}~\bibnamefont {Fujii}}, \bibinfo {author} {\bibfnamefont {J.~R.}\ \bibnamefont {McClean}}, \bibinfo {author} {\bibfnamefont {K.}~\bibnamefont {Mitarai}}, \bibinfo {author} {\bibfnamefont {X.}~\bibnamefont {Yuan}}, \bibinfo {author} {\bibfnamefont {L.}~\bibnamefont {Cincio}},\ and\ \bibinfo {author} {\bibfnamefont {P.~J.}\ \bibnamefont {Coles}},\ }\href {https://doi.org/10.1038/s42254-021-00348-9} {\bibfield  {journal} {\bibinfo  {journal} {Nature Reviews Physics}\ }\textbf {\bibinfo {volume} {3}},\ \bibinfo {pages} {625} (\bibinfo {year} {2021})}\BibitemShut {NoStop}%
\bibitem [{\citenamefont {Messiah}(2014)}]{messiahQuantum}%
  \BibitemOpen
  \bibfield  {author} {\bibinfo {author} {\bibfnamefont {A.}~\bibnamefont {Messiah}},\ }\href@noop {} {\emph {\bibinfo {title} {Quantum mechanics}}}\ (\bibinfo  {publisher} {Courier Corporation},\ \bibinfo {year} {2014})\BibitemShut {NoStop}%
\bibitem [{\citenamefont {Clements}\ \emph {et~al.}(2016)\citenamefont {Clements}, \citenamefont {Humphreys}, \citenamefont {Metcalf}, \citenamefont {Kolthammer},\ and\ \citenamefont {Walmsley}}]{Clements:16}%
  \BibitemOpen
  \bibfield  {author} {\bibinfo {author} {\bibfnamefont {W.~R.}\ \bibnamefont {Clements}}, \bibinfo {author} {\bibfnamefont {P.~C.}\ \bibnamefont {Humphreys}}, \bibinfo {author} {\bibfnamefont {B.~J.}\ \bibnamefont {Metcalf}}, \bibinfo {author} {\bibfnamefont {W.~S.}\ \bibnamefont {Kolthammer}},\ and\ \bibinfo {author} {\bibfnamefont {I.~A.}\ \bibnamefont {Walmsley}},\ }\href {https://doi.org/10.1364/OPTICA.3.001460} {\bibfield  {journal} {\bibinfo  {journal} {Optica}\ }\textbf {\bibinfo {volume} {3}},\ \bibinfo {pages} {1460} (\bibinfo {year} {2016})}\BibitemShut {NoStop}%
\bibitem [{\citenamefont {Blekos}\ \emph {et~al.}(2023)\citenamefont {Blekos}, \citenamefont {Brand}, \citenamefont {Ceschini}, \citenamefont {Chou}, \citenamefont {Li}, \citenamefont {Pandya},\ and\ \citenamefont {Summer}}]{blekos2023review}%
  \BibitemOpen
  \bibfield  {author} {\bibinfo {author} {\bibfnamefont {K.}~\bibnamefont {Blekos}}, \bibinfo {author} {\bibfnamefont {D.}~\bibnamefont {Brand}}, \bibinfo {author} {\bibfnamefont {A.}~\bibnamefont {Ceschini}}, \bibinfo {author} {\bibfnamefont {C.-H.}\ \bibnamefont {Chou}}, \bibinfo {author} {\bibfnamefont {R.-H.}\ \bibnamefont {Li}}, \bibinfo {author} {\bibfnamefont {K.}~\bibnamefont {Pandya}},\ and\ \bibinfo {author} {\bibfnamefont {A.}~\bibnamefont {Summer}},\ }\bibfield  {journal} {\bibinfo  {journal} {arXiv preprint arXiv:2306.09198}\ }\href {https://doi.org/https://doi.org/10.48550/arXiv.2306.09198} {https://doi.org/10.48550/arXiv.2306.09198} (\bibinfo {year} {2023})\BibitemShut {NoStop}%
\bibitem [{\citenamefont {Tilly}\ \emph {et~al.}(2022)\citenamefont {Tilly}, \citenamefont {Chen}, \citenamefont {Cao}, \citenamefont {Picozzi}, \citenamefont {Setia}, \citenamefont {Li}, \citenamefont {Grant}, \citenamefont {Wossnig}, \citenamefont {Rungger}, \citenamefont {Booth},\ and\ \citenamefont {Tennyson}}]{TILLY20221}%
  \BibitemOpen
  \bibfield  {author} {\bibinfo {author} {\bibfnamefont {J.}~\bibnamefont {Tilly}}, \bibinfo {author} {\bibfnamefont {H.}~\bibnamefont {Chen}}, \bibinfo {author} {\bibfnamefont {S.}~\bibnamefont {Cao}}, \bibinfo {author} {\bibfnamefont {D.}~\bibnamefont {Picozzi}}, \bibinfo {author} {\bibfnamefont {K.}~\bibnamefont {Setia}}, \bibinfo {author} {\bibfnamefont {Y.}~\bibnamefont {Li}}, \bibinfo {author} {\bibfnamefont {E.}~\bibnamefont {Grant}}, \bibinfo {author} {\bibfnamefont {L.}~\bibnamefont {Wossnig}}, \bibinfo {author} {\bibfnamefont {I.}~\bibnamefont {Rungger}}, \bibinfo {author} {\bibfnamefont {G.~H.}\ \bibnamefont {Booth}},\ and\ \bibinfo {author} {\bibfnamefont {J.}~\bibnamefont {Tennyson}},\ }\href {https://doi.org/https://doi.org/10.1016/j.physrep.2022.08.003} {\bibfield  {journal} {\bibinfo  {journal} {Physics Reports}\ }\textbf {\bibinfo {volume} {986}},\ \bibinfo {pages} {1} (\bibinfo {year} {2022})},\ \bibinfo {note} {the Variational Quantum Eigensolver: a review of methods and best
  practices}\BibitemShut {NoStop}%
\bibitem [{\citenamefont {Cao}\ \emph {et~al.}(2019)\citenamefont {Cao}, \citenamefont {Romero}, \citenamefont {Olson}, \citenamefont {Degroote}, \citenamefont {Johnson}, \citenamefont {Kieferová}, \citenamefont {Kivlichan}, \citenamefont {Menke}, \citenamefont {Peropadre}, \citenamefont {Sawaya}, \citenamefont {Sim}, \citenamefont {Veis},\ and\ \citenamefont {Aspuru-Guzik}}]{cao_quantum_2019}%
  \BibitemOpen
  \bibfield  {author} {\bibinfo {author} {\bibfnamefont {Y.}~\bibnamefont {Cao}}, \bibinfo {author} {\bibfnamefont {J.}~\bibnamefont {Romero}}, \bibinfo {author} {\bibfnamefont {J.~P.}\ \bibnamefont {Olson}}, \bibinfo {author} {\bibfnamefont {M.}~\bibnamefont {Degroote}}, \bibinfo {author} {\bibfnamefont {P.~D.}\ \bibnamefont {Johnson}}, \bibinfo {author} {\bibfnamefont {M.}~\bibnamefont {Kieferová}}, \bibinfo {author} {\bibfnamefont {I.~D.}\ \bibnamefont {Kivlichan}}, \bibinfo {author} {\bibfnamefont {T.}~\bibnamefont {Menke}}, \bibinfo {author} {\bibfnamefont {B.}~\bibnamefont {Peropadre}}, \bibinfo {author} {\bibfnamefont {N.~P.~D.}\ \bibnamefont {Sawaya}}, \bibinfo {author} {\bibfnamefont {S.}~\bibnamefont {Sim}}, \bibinfo {author} {\bibfnamefont {L.}~\bibnamefont {Veis}},\ and\ \bibinfo {author} {\bibfnamefont {A.}~\bibnamefont {Aspuru-Guzik}},\ }\href {https://doi.org/10.1021/acs.chemrev.8b00803} {\bibfield  {journal} {\bibinfo  {journal} {Chemical Reviews}\ }\textbf {\bibinfo {volume} {119}},\
  \bibinfo {pages} {10856} (\bibinfo {year} {2019})},\ \bibinfo {note} {publisher: American Chemical Society}\BibitemShut {NoStop}%
\bibitem [{\citenamefont {Brandhofer}\ \emph {et~al.}(2022)\citenamefont {Brandhofer}, \citenamefont {Braun}, \citenamefont {Dehn}, \citenamefont {Hellstern}, \citenamefont {Hüls}, \citenamefont {Ji}, \citenamefont {Polian}, \citenamefont {Bhatia},\ and\ \citenamefont {Wellens}}]{brandhofer_benchmarking_2022}%
  \BibitemOpen
  \bibfield  {author} {\bibinfo {author} {\bibfnamefont {S.}~\bibnamefont {Brandhofer}}, \bibinfo {author} {\bibfnamefont {D.}~\bibnamefont {Braun}}, \bibinfo {author} {\bibfnamefont {V.}~\bibnamefont {Dehn}}, \bibinfo {author} {\bibfnamefont {G.}~\bibnamefont {Hellstern}}, \bibinfo {author} {\bibfnamefont {M.}~\bibnamefont {Hüls}}, \bibinfo {author} {\bibfnamefont {Y.}~\bibnamefont {Ji}}, \bibinfo {author} {\bibfnamefont {I.}~\bibnamefont {Polian}}, \bibinfo {author} {\bibfnamefont {A.~S.}\ \bibnamefont {Bhatia}},\ and\ \bibinfo {author} {\bibfnamefont {T.}~\bibnamefont {Wellens}},\ }\href {https://doi.org/10.1007/s11128-022-03766-5} {\bibfield  {journal} {\bibinfo  {journal} {Quantum Information Processing}\ }\textbf {\bibinfo {volume} {22}},\ \bibinfo {pages} {25} (\bibinfo {year} {2022})}\BibitemShut {NoStop}%
\bibitem [{\citenamefont {Barkoutsos}\ \emph {et~al.}(2020)\citenamefont {Barkoutsos}, \citenamefont {Nannicini}, \citenamefont {Robert}, \citenamefont {Tavernelli},\ and\ \citenamefont {Woerner}}]{Barkoutsos2020improving}%
  \BibitemOpen
  \bibfield  {author} {\bibinfo {author} {\bibfnamefont {P.~K.}\ \bibnamefont {Barkoutsos}}, \bibinfo {author} {\bibfnamefont {G.}~\bibnamefont {Nannicini}}, \bibinfo {author} {\bibfnamefont {A.}~\bibnamefont {Robert}}, \bibinfo {author} {\bibfnamefont {I.}~\bibnamefont {Tavernelli}},\ and\ \bibinfo {author} {\bibfnamefont {S.}~\bibnamefont {Woerner}},\ }\href {https://doi.org/10.22331/q-2020-04-20-256} {\bibfield  {journal} {\bibinfo  {journal} {{Quantum}}\ }\textbf {\bibinfo {volume} {4}},\ \bibinfo {pages} {256} (\bibinfo {year} {2020})}\BibitemShut {NoStop}%
\bibitem [{\citenamefont {Fernández-Lorenzo}\ \emph {et~al.}(2021)\citenamefont {Fernández-Lorenzo}, \citenamefont {Porras},\ and\ \citenamefont {García-Ripoll}}]{Fernández-Lorenzo_2021}%
  \BibitemOpen
  \bibfield  {author} {\bibinfo {author} {\bibfnamefont {S.}~\bibnamefont {Fernández-Lorenzo}}, \bibinfo {author} {\bibfnamefont {D.}~\bibnamefont {Porras}},\ and\ \bibinfo {author} {\bibfnamefont {J.~J.}\ \bibnamefont {García-Ripoll}},\ }\href {https://doi.org/10.1088/2058-9565/abf9af} {\bibfield  {journal} {\bibinfo  {journal} {Quantum Science and Technology}\ }\textbf {\bibinfo {volume} {6}},\ \bibinfo {pages} {034010} (\bibinfo {year} {2021})}\BibitemShut {NoStop}%
\bibitem [{\citenamefont {Hegade}\ \emph {et~al.}(2022)\citenamefont {Hegade}, \citenamefont {Chandarana}, \citenamefont {Paul}, \citenamefont {Chen}, \citenamefont {Albarr\'an-Arriagada},\ and\ \citenamefont {Solano}}]{Hegade2022portfolio}%
  \BibitemOpen
  \bibfield  {author} {\bibinfo {author} {\bibfnamefont {N.~N.}\ \bibnamefont {Hegade}}, \bibinfo {author} {\bibfnamefont {P.}~\bibnamefont {Chandarana}}, \bibinfo {author} {\bibfnamefont {K.}~\bibnamefont {Paul}}, \bibinfo {author} {\bibfnamefont {X.}~\bibnamefont {Chen}}, \bibinfo {author} {\bibfnamefont {F.}~\bibnamefont {Albarr\'an-Arriagada}},\ and\ \bibinfo {author} {\bibfnamefont {E.}~\bibnamefont {Solano}},\ }\href {https://doi.org/10.1103/PhysRevResearch.4.043204} {\bibfield  {journal} {\bibinfo  {journal} {Phys. Rev. Res.}\ }\textbf {\bibinfo {volume} {4}},\ \bibinfo {pages} {043204} (\bibinfo {year} {2022})}\BibitemShut {NoStop}%
\bibitem [{\citenamefont {Zhu}\ \emph {et~al.}(2022)\citenamefont {Zhu}, \citenamefont {Tang}, \citenamefont {Barron}, \citenamefont {Calderon-Vargas}, \citenamefont {Mayhall}, \citenamefont {Barnes},\ and\ \citenamefont {Economou}}]{Zhu2022AQAOA}%
  \BibitemOpen
  \bibfield  {author} {\bibinfo {author} {\bibfnamefont {L.}~\bibnamefont {Zhu}}, \bibinfo {author} {\bibfnamefont {H.~L.}\ \bibnamefont {Tang}}, \bibinfo {author} {\bibfnamefont {G.~S.}\ \bibnamefont {Barron}}, \bibinfo {author} {\bibfnamefont {F.~A.}\ \bibnamefont {Calderon-Vargas}}, \bibinfo {author} {\bibfnamefont {N.~J.}\ \bibnamefont {Mayhall}}, \bibinfo {author} {\bibfnamefont {E.}~\bibnamefont {Barnes}},\ and\ \bibinfo {author} {\bibfnamefont {S.~E.}\ \bibnamefont {Economou}},\ }\href {https://doi.org/10.1103/PhysRevResearch.4.033029} {\bibfield  {journal} {\bibinfo  {journal} {Phys. Rev. Res.}\ }\textbf {\bibinfo {volume} {4}},\ \bibinfo {pages} {033029} (\bibinfo {year} {2022})}\BibitemShut {NoStop}%
\bibitem [{\citenamefont {Oh}\ \emph {et~al.}(2019)\citenamefont {Oh}, \citenamefont {Mohammadbagherpoor}, \citenamefont {Dreher}, \citenamefont {Singh}, \citenamefont {Yu},\ and\ \citenamefont {Rindos}}]{oh2019solving}%
  \BibitemOpen
  \bibfield  {author} {\bibinfo {author} {\bibfnamefont {Y.-H.}\ \bibnamefont {Oh}}, \bibinfo {author} {\bibfnamefont {H.}~\bibnamefont {Mohammadbagherpoor}}, \bibinfo {author} {\bibfnamefont {P.}~\bibnamefont {Dreher}}, \bibinfo {author} {\bibfnamefont {A.}~\bibnamefont {Singh}}, \bibinfo {author} {\bibfnamefont {X.}~\bibnamefont {Yu}},\ and\ \bibinfo {author} {\bibfnamefont {A.~J.}\ \bibnamefont {Rindos}},\ }\bibfield  {journal} {\bibinfo  {journal} {arXiv preprint arXiv:1911.00595}\ }\href {https://doi.org/https://doi.org/10.48550/arXiv.1911.00595} {https://doi.org/10.48550/arXiv.1911.00595} (\bibinfo {year} {2019})\BibitemShut {NoStop}%
\bibitem [{\citenamefont {Zhou}\ \emph {et~al.}(2023)\citenamefont {Zhou}, \citenamefont {Du}, \citenamefont {Tian},\ and\ \citenamefont {Tao}}]{PhysRevApplied.19.024027}%
  \BibitemOpen
  \bibfield  {author} {\bibinfo {author} {\bibfnamefont {Z.}~\bibnamefont {Zhou}}, \bibinfo {author} {\bibfnamefont {Y.}~\bibnamefont {Du}}, \bibinfo {author} {\bibfnamefont {X.}~\bibnamefont {Tian}},\ and\ \bibinfo {author} {\bibfnamefont {D.}~\bibnamefont {Tao}},\ }\href {https://doi.org/10.1103/PhysRevApplied.19.024027} {\bibfield  {journal} {\bibinfo  {journal} {Phys. Rev. Appl.}\ }\textbf {\bibinfo {volume} {19}},\ \bibinfo {pages} {024027} (\bibinfo {year} {2023})}\BibitemShut {NoStop}%
\bibitem [{\citenamefont {Shor}(1994)}]{shor1994algorithms}%
  \BibitemOpen
  \bibfield  {author} {\bibinfo {author} {\bibfnamefont {P.}~\bibnamefont {Shor}},\ }\href {https://doi.org/10.1109/SFCS.1994.365700} {\bibfield  {journal} {\bibinfo  {journal} {Proceedings 35th Annual Symposium on Foundations of Computer Science, IEEE}\ ,\ \bibinfo {pages} {124}} (\bibinfo {year} {1994})}\BibitemShut {NoStop}%
\bibitem [{\citenamefont {Jiang}\ \emph {et~al.}(2018)\citenamefont {Jiang}, \citenamefont {Britt}, \citenamefont {McCaskey}, \citenamefont {Humble},\ and\ \citenamefont {Kais}}]{jiang2018quantum}%
  \BibitemOpen
  \bibfield  {author} {\bibinfo {author} {\bibfnamefont {S.}~\bibnamefont {Jiang}}, \bibinfo {author} {\bibfnamefont {K.~A.}\ \bibnamefont {Britt}}, \bibinfo {author} {\bibfnamefont {A.~J.}\ \bibnamefont {McCaskey}}, \bibinfo {author} {\bibfnamefont {T.~S.}\ \bibnamefont {Humble}},\ and\ \bibinfo {author} {\bibfnamefont {S.}~\bibnamefont {Kais}},\ }\href {https://www.nature.com/articles/srep43048} {\bibfield  {journal} {\bibinfo  {journal} {Scientific reports}\ }\textbf {\bibinfo {volume} {8}},\ \bibinfo {pages} {1} (\bibinfo {year} {2018})}\BibitemShut {NoStop}%
\bibitem [{\citenamefont {Peng}\ \emph {et~al.}(2008{\natexlab{a}})\citenamefont {Peng}, \citenamefont {Liao}, \citenamefont {Xu}, \citenamefont {Qin}, \citenamefont {Zhou}, \citenamefont {Suter},\ and\ \citenamefont {Du}}]{Peng2008FactorPrl}%
  \BibitemOpen
  \bibfield  {author} {\bibinfo {author} {\bibfnamefont {X.}~\bibnamefont {Peng}}, \bibinfo {author} {\bibfnamefont {Z.}~\bibnamefont {Liao}}, \bibinfo {author} {\bibfnamefont {N.}~\bibnamefont {Xu}}, \bibinfo {author} {\bibfnamefont {G.}~\bibnamefont {Qin}}, \bibinfo {author} {\bibfnamefont {X.}~\bibnamefont {Zhou}}, \bibinfo {author} {\bibfnamefont {D.}~\bibnamefont {Suter}},\ and\ \bibinfo {author} {\bibfnamefont {J.}~\bibnamefont {Du}},\ }\href {https://doi.org/10.1103/PhysRevLett.101.220405} {\bibfield  {journal} {\bibinfo  {journal} {Phys. Rev. Lett.}\ }\textbf {\bibinfo {volume} {101}},\ \bibinfo {pages} {220405} (\bibinfo {year} {2008}{\natexlab{a}})}\BibitemShut {NoStop}%
\bibitem [{\citenamefont {Hegade}\ \emph {et~al.}(2021)\citenamefont {Hegade}, \citenamefont {Paul}, \citenamefont {Albarr\'an-Arriagada}, \citenamefont {Chen},\ and\ \citenamefont {Solano}}]{Hegade2021Fac}%
  \BibitemOpen
  \bibfield  {author} {\bibinfo {author} {\bibfnamefont {N.~N.}\ \bibnamefont {Hegade}}, \bibinfo {author} {\bibfnamefont {K.}~\bibnamefont {Paul}}, \bibinfo {author} {\bibfnamefont {F.}~\bibnamefont {Albarr\'an-Arriagada}}, \bibinfo {author} {\bibfnamefont {X.}~\bibnamefont {Chen}},\ and\ \bibinfo {author} {\bibfnamefont {E.}~\bibnamefont {Solano}},\ }\href {https://doi.org/10.1103/PhysRevA.104.L050403} {\bibfield  {journal} {\bibinfo  {journal} {Phys. Rev. A}\ }\textbf {\bibinfo {volume} {104}},\ \bibinfo {pages} {L050403} (\bibinfo {year} {2021})}\BibitemShut {NoStop}%
\bibitem [{\citenamefont {Karamlou}\ \emph {et~al.}(2021)\citenamefont {Karamlou}, \citenamefont {Simon}, \citenamefont {Katabarwa}, \citenamefont {Scholten}, \citenamefont {Peropadre},\ and\ \citenamefont {Cao}}]{karamlou2021analyzing}%
  \BibitemOpen
  \bibfield  {author} {\bibinfo {author} {\bibfnamefont {A.~H.}\ \bibnamefont {Karamlou}}, \bibinfo {author} {\bibfnamefont {W.~A.}\ \bibnamefont {Simon}}, \bibinfo {author} {\bibfnamefont {A.}~\bibnamefont {Katabarwa}}, \bibinfo {author} {\bibfnamefont {T.~L.}\ \bibnamefont {Scholten}}, \bibinfo {author} {\bibfnamefont {B.}~\bibnamefont {Peropadre}},\ and\ \bibinfo {author} {\bibfnamefont {Y.}~\bibnamefont {Cao}},\ }\href {https://doi.org/10.1038/s41534-021-00478-z} {\bibfield  {journal} {\bibinfo  {journal} {npj Quantum Information}\ }\textbf {\bibinfo {volume} {7}},\ \bibinfo {pages} {156} (\bibinfo {year} {2021})}\BibitemShut {NoStop}%
\bibitem [{\citenamefont {Yan}\ \emph {et~al.}(2022)\citenamefont {Yan}, \citenamefont {Tan}, \citenamefont {Wei}, \citenamefont {Jiang}, \citenamefont {Wang}, \citenamefont {Wang}, \citenamefont {Luo}, \citenamefont {Duan}, \citenamefont {Liu}, \citenamefont {Shi} \emph {et~al.}}]{yan2022factoring}%
  \BibitemOpen
  \bibfield  {author} {\bibinfo {author} {\bibfnamefont {B.}~\bibnamefont {Yan}}, \bibinfo {author} {\bibfnamefont {Z.}~\bibnamefont {Tan}}, \bibinfo {author} {\bibfnamefont {S.}~\bibnamefont {Wei}}, \bibinfo {author} {\bibfnamefont {H.}~\bibnamefont {Jiang}}, \bibinfo {author} {\bibfnamefont {W.}~\bibnamefont {Wang}}, \bibinfo {author} {\bibfnamefont {H.}~\bibnamefont {Wang}}, \bibinfo {author} {\bibfnamefont {L.}~\bibnamefont {Luo}}, \bibinfo {author} {\bibfnamefont {Q.}~\bibnamefont {Duan}}, \bibinfo {author} {\bibfnamefont {Y.}~\bibnamefont {Liu}}, \bibinfo {author} {\bibfnamefont {W.}~\bibnamefont {Shi}}, \emph {et~al.},\ }\bibfield  {journal} {\bibinfo  {journal} {arXiv preprint arXiv:2212.12372}\ }\href {https://doi.org/10.48550/arXiv.2212.12372} {10.48550/arXiv.2212.12372} (\bibinfo {year} {2022})\BibitemShut {NoStop}%
\bibitem [{\citenamefont {Peng}\ \emph {et~al.}(2008{\natexlab{b}})\citenamefont {Peng}, \citenamefont {Liao}, \citenamefont {Xu}, \citenamefont {Qin}, \citenamefont {Zhou}, \citenamefont {Suter},\ and\ \citenamefont {Du}}]{peng2008quantum}%
  \BibitemOpen
  \bibfield  {author} {\bibinfo {author} {\bibfnamefont {X.}~\bibnamefont {Peng}}, \bibinfo {author} {\bibfnamefont {Z.}~\bibnamefont {Liao}}, \bibinfo {author} {\bibfnamefont {N.}~\bibnamefont {Xu}}, \bibinfo {author} {\bibfnamefont {G.}~\bibnamefont {Qin}}, \bibinfo {author} {\bibfnamefont {X.}~\bibnamefont {Zhou}}, \bibinfo {author} {\bibfnamefont {D.}~\bibnamefont {Suter}},\ and\ \bibinfo {author} {\bibfnamefont {J.}~\bibnamefont {Du}},\ }\href {https://journals.aps.org/prl/abstract/10.1103/PhysRevLett.101.220405} {\bibfield  {journal} {\bibinfo  {journal} {Physical review letters}\ }\textbf {\bibinfo {volume} {101}},\ \bibinfo {pages} {220405} (\bibinfo {year} {2008}{\natexlab{b}})}\BibitemShut {NoStop}%
\bibitem [{\citenamefont {Bharti}\ \emph {et~al.}(2022)\citenamefont {Bharti}, \citenamefont {Cervera-Lierta}, \citenamefont {Kyaw}, \citenamefont {Haug}, \citenamefont {Alperin-Lea}, \citenamefont {Anand}, \citenamefont {Degroote}, \citenamefont {Heimonen}, \citenamefont {Kottmann}, \citenamefont {Menke}, \citenamefont {Mok}, \citenamefont {Sim}, \citenamefont {Kwek},\ and\ \citenamefont {Aspuru-Guzik}}]{Bharati2002review}%
  \BibitemOpen
  \bibfield  {author} {\bibinfo {author} {\bibfnamefont {K.}~\bibnamefont {Bharti}}, \bibinfo {author} {\bibfnamefont {A.}~\bibnamefont {Cervera-Lierta}}, \bibinfo {author} {\bibfnamefont {T.~H.}\ \bibnamefont {Kyaw}}, \bibinfo {author} {\bibfnamefont {T.}~\bibnamefont {Haug}}, \bibinfo {author} {\bibfnamefont {S.}~\bibnamefont {Alperin-Lea}}, \bibinfo {author} {\bibfnamefont {A.}~\bibnamefont {Anand}}, \bibinfo {author} {\bibfnamefont {M.}~\bibnamefont {Degroote}}, \bibinfo {author} {\bibfnamefont {H.}~\bibnamefont {Heimonen}}, \bibinfo {author} {\bibfnamefont {J.~S.}\ \bibnamefont {Kottmann}}, \bibinfo {author} {\bibfnamefont {T.}~\bibnamefont {Menke}}, \bibinfo {author} {\bibfnamefont {W.-K.}\ \bibnamefont {Mok}}, \bibinfo {author} {\bibfnamefont {S.}~\bibnamefont {Sim}}, \bibinfo {author} {\bibfnamefont {L.-C.}\ \bibnamefont {Kwek}},\ and\ \bibinfo {author} {\bibfnamefont {A.}~\bibnamefont {Aspuru-Guzik}},\ }\href {https://doi.org/10.1103/RevModPhys.94.015004} {\bibfield  {journal} {\bibinfo  {journal}
  {Rev. Mod. Phys.}\ }\textbf {\bibinfo {volume} {94}},\ \bibinfo {pages} {015004} (\bibinfo {year} {2022})}\BibitemShut {NoStop}%
\bibitem [{\citenamefont {Zhong}\ \emph {et~al.}(2020)\citenamefont {Zhong}, \citenamefont {Wang}, \citenamefont {Deng}, \citenamefont {Chen}, \citenamefont {Peng}, \citenamefont {Luo}, \citenamefont {Qin}, \citenamefont {Wu}, \citenamefont {Ding}, \citenamefont {Hu} \emph {et~al.}}]{zhong2020quantum}%
  \BibitemOpen
  \bibfield  {author} {\bibinfo {author} {\bibfnamefont {H.-S.}\ \bibnamefont {Zhong}}, \bibinfo {author} {\bibfnamefont {H.}~\bibnamefont {Wang}}, \bibinfo {author} {\bibfnamefont {Y.-H.}\ \bibnamefont {Deng}}, \bibinfo {author} {\bibfnamefont {M.-C.}\ \bibnamefont {Chen}}, \bibinfo {author} {\bibfnamefont {L.-C.}\ \bibnamefont {Peng}}, \bibinfo {author} {\bibfnamefont {Y.-H.}\ \bibnamefont {Luo}}, \bibinfo {author} {\bibfnamefont {J.}~\bibnamefont {Qin}}, \bibinfo {author} {\bibfnamefont {D.}~\bibnamefont {Wu}}, \bibinfo {author} {\bibfnamefont {X.}~\bibnamefont {Ding}}, \bibinfo {author} {\bibfnamefont {Y.}~\bibnamefont {Hu}}, \emph {et~al.},\ }\href {https://doi.org/10.1126/science.abe8770} {\bibfield  {journal} {\bibinfo  {journal} {Science}\ }\textbf {\bibinfo {volume} {370}},\ \bibinfo {pages} {1460} (\bibinfo {year} {2020})}\BibitemShut {NoStop}%
\bibitem [{\citenamefont {Madsen}\ \emph {et~al.}(2022)\citenamefont {Madsen}, \citenamefont {Laudenbach}, \citenamefont {Askarani}, \citenamefont {Rortais}, \citenamefont {Vincent}, \citenamefont {Bulmer}, \citenamefont {Miatto}, \citenamefont {Neuhaus}, \citenamefont {Helt}, \citenamefont {Collins} \emph {et~al.}}]{madsen2022quantum}%
  \BibitemOpen
  \bibfield  {author} {\bibinfo {author} {\bibfnamefont {L.~S.}\ \bibnamefont {Madsen}}, \bibinfo {author} {\bibfnamefont {F.}~\bibnamefont {Laudenbach}}, \bibinfo {author} {\bibfnamefont {M.~F.}\ \bibnamefont {Askarani}}, \bibinfo {author} {\bibfnamefont {F.}~\bibnamefont {Rortais}}, \bibinfo {author} {\bibfnamefont {T.}~\bibnamefont {Vincent}}, \bibinfo {author} {\bibfnamefont {J.~F.}\ \bibnamefont {Bulmer}}, \bibinfo {author} {\bibfnamefont {F.~M.}\ \bibnamefont {Miatto}}, \bibinfo {author} {\bibfnamefont {L.}~\bibnamefont {Neuhaus}}, \bibinfo {author} {\bibfnamefont {L.~G.}\ \bibnamefont {Helt}}, \bibinfo {author} {\bibfnamefont {M.~J.}\ \bibnamefont {Collins}}, \emph {et~al.},\ }\href {https://doi.org/https://doi.org/10.1038/s41586-022-04725-x} {\bibfield  {journal} {\bibinfo  {journal} {Nature}\ }\textbf {\bibinfo {volume} {606}},\ \bibinfo {pages} {75} (\bibinfo {year} {2022})}\BibitemShut {NoStop}%
\bibitem [{\citenamefont {Peruzzo}\ \emph {et~al.}(2014)\citenamefont {Peruzzo}, \citenamefont {McClean}, \citenamefont {Shadbolt}, \citenamefont {Yung}, \citenamefont {Zhou}, \citenamefont {Love}, \citenamefont {Aspuru-Guzik},\ and\ \citenamefont {O’brien}}]{peruzzo2014variational}%
  \BibitemOpen
  \bibfield  {author} {\bibinfo {author} {\bibfnamefont {A.}~\bibnamefont {Peruzzo}}, \bibinfo {author} {\bibfnamefont {J.}~\bibnamefont {McClean}}, \bibinfo {author} {\bibfnamefont {P.}~\bibnamefont {Shadbolt}}, \bibinfo {author} {\bibfnamefont {M.-H.}\ \bibnamefont {Yung}}, \bibinfo {author} {\bibfnamefont {X.-Q.}\ \bibnamefont {Zhou}}, \bibinfo {author} {\bibfnamefont {P.~J.}\ \bibnamefont {Love}}, \bibinfo {author} {\bibfnamefont {A.}~\bibnamefont {Aspuru-Guzik}},\ and\ \bibinfo {author} {\bibfnamefont {J.~L.}\ \bibnamefont {O’brien}},\ }\href {https://doi.org/https://doi.org/10.1038/ncomms5213} {\bibfield  {journal} {\bibinfo  {journal} {Nature communications}\ }\textbf {\bibinfo {volume} {5}},\ \bibinfo {pages} {4213} (\bibinfo {year} {2014})}\BibitemShut {NoStop}%
\bibitem [{\citenamefont {Lee}\ \emph {et~al.}(2022)\citenamefont {Lee}, \citenamefont {Lee}, \citenamefont {Hong}, \citenamefont {Lim}, \citenamefont {Cho}, \citenamefont {Han}, \citenamefont {Shin}, \citenamefont {ur~Rehman},\ and\ \citenamefont {Kim}}]{lee2022error}%
  \BibitemOpen
  \bibfield  {author} {\bibinfo {author} {\bibfnamefont {D.}~\bibnamefont {Lee}}, \bibinfo {author} {\bibfnamefont {J.}~\bibnamefont {Lee}}, \bibinfo {author} {\bibfnamefont {S.}~\bibnamefont {Hong}}, \bibinfo {author} {\bibfnamefont {H.-T.}\ \bibnamefont {Lim}}, \bibinfo {author} {\bibfnamefont {Y.-W.}\ \bibnamefont {Cho}}, \bibinfo {author} {\bibfnamefont {S.-W.}\ \bibnamefont {Han}}, \bibinfo {author} {\bibfnamefont {H.}~\bibnamefont {Shin}}, \bibinfo {author} {\bibfnamefont {J.}~\bibnamefont {ur~Rehman}},\ and\ \bibinfo {author} {\bibfnamefont {Y.-S.}\ \bibnamefont {Kim}},\ }\href {https://doi.org/https://doi.org/10.1364/OPTICA.441163} {\bibfield  {journal} {\bibinfo  {journal} {Optica}\ }\textbf {\bibinfo {volume} {9}},\ \bibinfo {pages} {88} (\bibinfo {year} {2022})}\BibitemShut {NoStop}%
\bibitem [{\citenamefont {Cimini}\ \emph {et~al.}(2024)\citenamefont {Cimini}, \citenamefont {Valeri}, \citenamefont {Piacentini}, \citenamefont {Ceccarelli}, \citenamefont {Corrielli}, \citenamefont {Osellame}, \citenamefont {Spagnolo},\ and\ \citenamefont {Sciarrino}}]{cimini2024variational}%
  \BibitemOpen
  \bibfield  {author} {\bibinfo {author} {\bibfnamefont {V.}~\bibnamefont {Cimini}}, \bibinfo {author} {\bibfnamefont {M.}~\bibnamefont {Valeri}}, \bibinfo {author} {\bibfnamefont {S.}~\bibnamefont {Piacentini}}, \bibinfo {author} {\bibfnamefont {F.}~\bibnamefont {Ceccarelli}}, \bibinfo {author} {\bibfnamefont {G.}~\bibnamefont {Corrielli}}, \bibinfo {author} {\bibfnamefont {R.}~\bibnamefont {Osellame}}, \bibinfo {author} {\bibfnamefont {N.}~\bibnamefont {Spagnolo}},\ and\ \bibinfo {author} {\bibfnamefont {F.}~\bibnamefont {Sciarrino}},\ }\href {https://doi.org/https://doi.org/10.1038/s41534-024-00821-0} {\bibfield  {journal} {\bibinfo  {journal} {npj Quantum Information}\ }\textbf {\bibinfo {volume} {10}},\ \bibinfo {pages} {26} (\bibinfo {year} {2024})}\BibitemShut {NoStop}%
\bibitem [{\citenamefont {Kandala}\ \emph {et~al.}(2017{\natexlab{a}})\citenamefont {Kandala}, \citenamefont {Mezzacapo}, \citenamefont {Temme}, \citenamefont {Takita}, \citenamefont {Brink}, \citenamefont {Chow},\ and\ \citenamefont {Gambetta}}]{KandalaHEA2017}%
  \BibitemOpen
  \bibfield  {author} {\bibinfo {author} {\bibfnamefont {A.}~\bibnamefont {Kandala}}, \bibinfo {author} {\bibfnamefont {A.}~\bibnamefont {Mezzacapo}}, \bibinfo {author} {\bibfnamefont {K.}~\bibnamefont {Temme}}, \bibinfo {author} {\bibfnamefont {M.}~\bibnamefont {Takita}}, \bibinfo {author} {\bibfnamefont {M.}~\bibnamefont {Brink}}, \bibinfo {author} {\bibfnamefont {J.~M.}\ \bibnamefont {Chow}},\ and\ \bibinfo {author} {\bibfnamefont {J.~M.}\ \bibnamefont {Gambetta}},\ }\href {https://doi.org/10.1038/nature23879} {\bibfield  {journal} {\bibinfo  {journal} {Nature}\ }\textbf {\bibinfo {volume} {549}},\ \bibinfo {pages} {242} (\bibinfo {year} {2017}{\natexlab{a}})}\BibitemShut {NoStop}%
\bibitem [{\citenamefont {Benedetti}\ \emph {et~al.}(2021)\citenamefont {Benedetti}, \citenamefont {Fiorentini},\ and\ \citenamefont {Lubasch}}]{Benedetti2021}%
  \BibitemOpen
  \bibfield  {author} {\bibinfo {author} {\bibfnamefont {M.}~\bibnamefont {Benedetti}}, \bibinfo {author} {\bibfnamefont {M.}~\bibnamefont {Fiorentini}},\ and\ \bibinfo {author} {\bibfnamefont {M.}~\bibnamefont {Lubasch}},\ }\href {https://doi.org/10.1103/PhysRevResearch.3.033083} {\bibfield  {journal} {\bibinfo  {journal} {Phys. Rev. Res.}\ }\textbf {\bibinfo {volume} {3}},\ \bibinfo {pages} {033083} (\bibinfo {year} {2021})}\BibitemShut {NoStop}%
\bibitem [{\citenamefont {Leone}\ \emph {et~al.}(2022)\citenamefont {Leone}, \citenamefont {Oliviero}, \citenamefont {Cincio},\ and\ \citenamefont {Cerezo}}]{leone2022practical}%
  \BibitemOpen
  \bibfield  {author} {\bibinfo {author} {\bibfnamefont {L.}~\bibnamefont {Leone}}, \bibinfo {author} {\bibfnamefont {S.~F.~E.}\ \bibnamefont {Oliviero}}, \bibinfo {author} {\bibfnamefont {L.}~\bibnamefont {Cincio}},\ and\ \bibinfo {author} {\bibfnamefont {M.}~\bibnamefont {Cerezo}},\ }\href {https://doi.org/https://doi.org/10.48550/arXiv.2211.01477} {\bibinfo {title} {On the practical usefulness of the hardware efficient ansatz}} (\bibinfo {year} {2022}),\ \Eprint {https://arxiv.org/abs/2211.01477} {arXiv:2211.01477 [quant-ph]} \BibitemShut {NoStop}%
\bibitem [{\citenamefont {Hardy}\ and\ \citenamefont {Wright}(1979)}]{hardy1979introduction}%
  \BibitemOpen
  \bibfield  {author} {\bibinfo {author} {\bibfnamefont {G.~H.}\ \bibnamefont {Hardy}}\ and\ \bibinfo {author} {\bibfnamefont {E.~M.}\ \bibnamefont {Wright}},\ }\href@noop {} {\emph {\bibinfo {title} {An introduction to the theory of numbers}}}\ (\bibinfo  {publisher} {Oxford university press},\ \bibinfo {year} {1979})\BibitemShut {NoStop}%
\bibitem [{\citenamefont {Amico}\ \emph {et~al.}(2019)\citenamefont {Amico}, \citenamefont {Saleem},\ and\ \citenamefont {Kumph}}]{amico2019experimental}%
  \BibitemOpen
  \bibfield  {author} {\bibinfo {author} {\bibfnamefont {M.}~\bibnamefont {Amico}}, \bibinfo {author} {\bibfnamefont {Z.~H.}\ \bibnamefont {Saleem}},\ and\ \bibinfo {author} {\bibfnamefont {M.}~\bibnamefont {Kumph}},\ }\href {https://doi.org/10.1103/PhysRevA.100.012305} {\bibfield  {journal} {\bibinfo  {journal} {Physical Review A}\ }\textbf {\bibinfo {volume} {100}},\ \bibinfo {pages} {012305} (\bibinfo {year} {2019})}\BibitemShut {NoStop}%
\bibitem [{\citenamefont {Corrielli}\ \emph {et~al.}(2021)\citenamefont {Corrielli}, \citenamefont {Crespi},\ and\ \citenamefont {Osellame}}]{corrielli2021femtosecond}%
  \BibitemOpen
  \bibfield  {author} {\bibinfo {author} {\bibfnamefont {G.}~\bibnamefont {Corrielli}}, \bibinfo {author} {\bibfnamefont {A.}~\bibnamefont {Crespi}},\ and\ \bibinfo {author} {\bibfnamefont {R.}~\bibnamefont {Osellame}},\ }\href {https://doi.org/https://doi.org/10.1515/nanoph-2021-0419} {\bibfield  {journal} {\bibinfo  {journal} {Nanophotonics}\ }\textbf {\bibinfo {volume} {10}},\ \bibinfo {pages} {3789} (\bibinfo {year} {2021})}\BibitemShut {NoStop}%
\bibitem [{\citenamefont {Pentangelo}\ \emph {et~al.}(2024)\citenamefont {Pentangelo}, \citenamefont {Di~Giano}, \citenamefont {Piacentini}, \citenamefont {Arpe}, \citenamefont {Ceccarelli}, \citenamefont {Crespi},\ and\ \citenamefont {Osellame}}]{pentangelo2024high}%
  \BibitemOpen
  \bibfield  {author} {\bibinfo {author} {\bibfnamefont {C.}~\bibnamefont {Pentangelo}}, \bibinfo {author} {\bibfnamefont {N.}~\bibnamefont {Di~Giano}}, \bibinfo {author} {\bibfnamefont {S.}~\bibnamefont {Piacentini}}, \bibinfo {author} {\bibfnamefont {R.}~\bibnamefont {Arpe}}, \bibinfo {author} {\bibfnamefont {F.}~\bibnamefont {Ceccarelli}}, \bibinfo {author} {\bibfnamefont {A.}~\bibnamefont {Crespi}},\ and\ \bibinfo {author} {\bibfnamefont {R.}~\bibnamefont {Osellame}},\ }\href {https://doi.org/https://doi.org/10.1515/nanoph-2023-0636} {\bibfield  {journal} {\bibinfo  {journal} {Nanophotonics}\ }\textbf {\bibinfo {volume} {13}},\ \bibinfo {pages} {2259} (\bibinfo {year} {2024})}\BibitemShut {NoStop}%
\bibitem [{\citenamefont {Skosana}\ and\ \citenamefont {Tame}(2021)}]{skosana2021demonstration}%
  \BibitemOpen
  \bibfield  {author} {\bibinfo {author} {\bibfnamefont {U.}~\bibnamefont {Skosana}}\ and\ \bibinfo {author} {\bibfnamefont {M.}~\bibnamefont {Tame}},\ }\href {https://doi.org/10.1038/s41598-021-95973-w} {\bibfield  {journal} {\bibinfo  {journal} {Scientific reports}\ }\textbf {\bibinfo {volume} {11}},\ \bibinfo {pages} {16599} (\bibinfo {year} {2021})}\BibitemShut {NoStop}%
\bibitem [{\citenamefont {Kandala}\ \emph {et~al.}(2017{\natexlab{b}})\citenamefont {Kandala}, \citenamefont {Mezzacapo}, \citenamefont {Temme}, \citenamefont {Takita}, \citenamefont {Brink}, \citenamefont {Chow},\ and\ \citenamefont {Gambetta}}]{kandala2017hardware}%
  \BibitemOpen
  \bibfield  {author} {\bibinfo {author} {\bibfnamefont {A.}~\bibnamefont {Kandala}}, \bibinfo {author} {\bibfnamefont {A.}~\bibnamefont {Mezzacapo}}, \bibinfo {author} {\bibfnamefont {K.}~\bibnamefont {Temme}}, \bibinfo {author} {\bibfnamefont {M.}~\bibnamefont {Takita}}, \bibinfo {author} {\bibfnamefont {M.}~\bibnamefont {Brink}}, \bibinfo {author} {\bibfnamefont {J.~M.}\ \bibnamefont {Chow}},\ and\ \bibinfo {author} {\bibfnamefont {J.~M.}\ \bibnamefont {Gambetta}},\ }\href@noop {} {\bibfield  {journal} {\bibinfo  {journal} {nature}\ }\textbf {\bibinfo {volume} {549}},\ \bibinfo {pages} {242} (\bibinfo {year} {2017}{\natexlab{b}})}\BibitemShut {NoStop}%
\bibitem [{\citenamefont {O’Malley}\ \emph {et~al.}(2016)\citenamefont {O’Malley}, \citenamefont {Babbush}, \citenamefont {Kivlichan}, \citenamefont {Romero}, \citenamefont {McClean}, \citenamefont {Barends}, \citenamefont {Kelly}, \citenamefont {Roushan}, \citenamefont {Tranter}, \citenamefont {Ding} \emph {et~al.}}]{o2016scalable}%
  \BibitemOpen
  \bibfield  {author} {\bibinfo {author} {\bibfnamefont {P.~J.}\ \bibnamefont {O’Malley}}, \bibinfo {author} {\bibfnamefont {R.}~\bibnamefont {Babbush}}, \bibinfo {author} {\bibfnamefont {I.~D.}\ \bibnamefont {Kivlichan}}, \bibinfo {author} {\bibfnamefont {J.}~\bibnamefont {Romero}}, \bibinfo {author} {\bibfnamefont {J.~R.}\ \bibnamefont {McClean}}, \bibinfo {author} {\bibfnamefont {R.}~\bibnamefont {Barends}}, \bibinfo {author} {\bibfnamefont {J.}~\bibnamefont {Kelly}}, \bibinfo {author} {\bibfnamefont {P.}~\bibnamefont {Roushan}}, \bibinfo {author} {\bibfnamefont {A.}~\bibnamefont {Tranter}}, \bibinfo {author} {\bibfnamefont {N.}~\bibnamefont {Ding}}, \emph {et~al.},\ }\href@noop {} {\bibfield  {journal} {\bibinfo  {journal} {Physical Review X}\ }\textbf {\bibinfo {volume} {6}},\ \bibinfo {pages} {031007} (\bibinfo {year} {2016})}\BibitemShut {NoStop}%
\bibitem [{\citenamefont {Farhi}\ \emph {et~al.}(2014)\citenamefont {Farhi}, \citenamefont {Goldstone},\ and\ \citenamefont {Gutmann}}]{farhi2014quantum}%
  \BibitemOpen
  \bibfield  {author} {\bibinfo {author} {\bibfnamefont {E.}~\bibnamefont {Farhi}}, \bibinfo {author} {\bibfnamefont {J.}~\bibnamefont {Goldstone}},\ and\ \bibinfo {author} {\bibfnamefont {S.}~\bibnamefont {Gutmann}},\ }\href@noop {} {\bibfield  {journal} {\bibinfo  {journal} {arXiv preprint arXiv:1411.4028}\ } (\bibinfo {year} {2014})}\BibitemShut {NoStop}%
\bibitem [{\citenamefont {Knill}\ \emph {et~al.}(2001)\citenamefont {Knill}, \citenamefont {Laflamme},\ and\ \citenamefont {Milburn}}]{knill2001scheme}%
  \BibitemOpen
  \bibfield  {author} {\bibinfo {author} {\bibfnamefont {E.}~\bibnamefont {Knill}}, \bibinfo {author} {\bibfnamefont {R.}~\bibnamefont {Laflamme}},\ and\ \bibinfo {author} {\bibfnamefont {G.~J.}\ \bibnamefont {Milburn}},\ }\href {https://doi.org/ttps://doi.org/10.1038/35051009} {\bibfield  {journal} {\bibinfo  {journal} {nature}\ }\textbf {\bibinfo {volume} {409}},\ \bibinfo {pages} {46} (\bibinfo {year} {2001})}\BibitemShut {NoStop}%
\bibitem [{\citenamefont {Innocenti}\ \emph {et~al.}(2023)\citenamefont {Innocenti}, \citenamefont {Lorenzo}, \citenamefont {Palmisano}, \citenamefont {Ferraro}, \citenamefont {Paternostro},\ and\ \citenamefont {Palma}}]{innocenti2023potential}%
  \BibitemOpen
  \bibfield  {author} {\bibinfo {author} {\bibfnamefont {L.}~\bibnamefont {Innocenti}}, \bibinfo {author} {\bibfnamefont {S.}~\bibnamefont {Lorenzo}}, \bibinfo {author} {\bibfnamefont {I.}~\bibnamefont {Palmisano}}, \bibinfo {author} {\bibfnamefont {A.}~\bibnamefont {Ferraro}}, \bibinfo {author} {\bibfnamefont {M.}~\bibnamefont {Paternostro}},\ and\ \bibinfo {author} {\bibfnamefont {G.~M.}\ \bibnamefont {Palma}},\ }\href {https://doi.org/https://doi.org/10.1038/s42005-023-01233-w} {\bibfield  {journal} {\bibinfo  {journal} {Communications Physics}\ }\textbf {\bibinfo {volume} {6}},\ \bibinfo {pages} {118} (\bibinfo {year} {2023})}\BibitemShut {NoStop}%
\bibitem [{\citenamefont {Spagnolo}\ \emph {et~al.}(2022)\citenamefont {Spagnolo}, \citenamefont {Morris}, \citenamefont {Piacentini}, \citenamefont {Antesberger}, \citenamefont {Massa}, \citenamefont {Crespi}, \citenamefont {Ceccarelli}, \citenamefont {Osellame},\ and\ \citenamefont {Walther}}]{spagnolo2022experimental}%
  \BibitemOpen
  \bibfield  {author} {\bibinfo {author} {\bibfnamefont {M.}~\bibnamefont {Spagnolo}}, \bibinfo {author} {\bibfnamefont {J.}~\bibnamefont {Morris}}, \bibinfo {author} {\bibfnamefont {S.}~\bibnamefont {Piacentini}}, \bibinfo {author} {\bibfnamefont {M.}~\bibnamefont {Antesberger}}, \bibinfo {author} {\bibfnamefont {F.}~\bibnamefont {Massa}}, \bibinfo {author} {\bibfnamefont {A.}~\bibnamefont {Crespi}}, \bibinfo {author} {\bibfnamefont {F.}~\bibnamefont {Ceccarelli}}, \bibinfo {author} {\bibfnamefont {R.}~\bibnamefont {Osellame}},\ and\ \bibinfo {author} {\bibfnamefont {P.}~\bibnamefont {Walther}},\ }\href {https://doi.org/https://doi.org/10.1038/s41566-022-00973-5} {\bibfield  {journal} {\bibinfo  {journal} {Nature Photonics}\ }\textbf {\bibinfo {volume} {16}},\ \bibinfo {pages} {318} (\bibinfo {year} {2022})}\BibitemShut {NoStop}%
\bibitem [{\citenamefont {Govia}\ \emph {et~al.}(2022)\citenamefont {Govia}, \citenamefont {Ribeill}, \citenamefont {Rowlands},\ and\ \citenamefont {Ohki}}]{govia2022nonlinear}%
  \BibitemOpen
  \bibfield  {author} {\bibinfo {author} {\bibfnamefont {L.}~\bibnamefont {Govia}}, \bibinfo {author} {\bibfnamefont {G.}~\bibnamefont {Ribeill}}, \bibinfo {author} {\bibfnamefont {G.}~\bibnamefont {Rowlands}},\ and\ \bibinfo {author} {\bibfnamefont {T.}~\bibnamefont {Ohki}},\ }\href {https://doi.org/10.1088/2634-4386/ac4fcd} {\bibfield  {journal} {\bibinfo  {journal} {Neuromorphic Computing and Engineering}\ }\textbf {\bibinfo {volume} {2}},\ \bibinfo {pages} {014008} (\bibinfo {year} {2022})}\BibitemShut {NoStop}%
\bibitem [{\citenamefont {Hamerly}\ \emph {et~al.}(2019)\citenamefont {Hamerly}, \citenamefont {Bernstein}, \citenamefont {Sludds}, \citenamefont {Solja{\v{c}}i{\'c}},\ and\ \citenamefont {Englund}}]{hamerly2019large}%
  \BibitemOpen
  \bibfield  {author} {\bibinfo {author} {\bibfnamefont {R.}~\bibnamefont {Hamerly}}, \bibinfo {author} {\bibfnamefont {L.}~\bibnamefont {Bernstein}}, \bibinfo {author} {\bibfnamefont {A.}~\bibnamefont {Sludds}}, \bibinfo {author} {\bibfnamefont {M.}~\bibnamefont {Solja{\v{c}}i{\'c}}},\ and\ \bibinfo {author} {\bibfnamefont {D.}~\bibnamefont {Englund}},\ }\href {https://doi.org/10.1103/PhysRevX.9.021032} {\bibfield  {journal} {\bibinfo  {journal} {Physical Review X}\ }\textbf {\bibinfo {volume} {9}},\ \bibinfo {pages} {021032} (\bibinfo {year} {2019})}\BibitemShut {NoStop}%
\bibitem [{\citenamefont {Mengu}\ \emph {et~al.}(2022)\citenamefont {Mengu}, \citenamefont {Rahman}, \citenamefont {Luo}, \citenamefont {Li}, \citenamefont {Kulce},\ and\ \citenamefont {Ozcan}}]{Mengu2022deeplearning}%
  \BibitemOpen
  \bibfield  {author} {\bibinfo {author} {\bibfnamefont {D.}~\bibnamefont {Mengu}}, \bibinfo {author} {\bibfnamefont {M.~S.~S.}\ \bibnamefont {Rahman}}, \bibinfo {author} {\bibfnamefont {Y.}~\bibnamefont {Luo}}, \bibinfo {author} {\bibfnamefont {J.}~\bibnamefont {Li}}, \bibinfo {author} {\bibfnamefont {O.}~\bibnamefont {Kulce}},\ and\ \bibinfo {author} {\bibfnamefont {A.}~\bibnamefont {Ozcan}},\ }\href {https://doi.org/10.1364/AOP.450345} {\bibfield  {journal} {\bibinfo  {journal} {Adv. Opt. Photon.}\ }\textbf {\bibinfo {volume} {14}},\ \bibinfo {pages} {209} (\bibinfo {year} {2022})}\BibitemShut {NoStop}%
\bibitem [{\citenamefont {Wang}\ \emph {et~al.}(2023)\citenamefont {Wang}, \citenamefont {Xue}, \citenamefont {Wang}, \citenamefont {Liu}, \citenamefont {Ding}, \citenamefont {Shi}, \citenamefont {Wang}, \citenamefont {Liu}, \citenamefont {Fu}, \citenamefont {Huang}, \citenamefont {Huang}, \citenamefont {Deng},\ and\ \citenamefont {Wu}}]{Wang23photonicLearning}%
  \BibitemOpen
  \bibfield  {author} {\bibinfo {author} {\bibfnamefont {Y.}~\bibnamefont {Wang}}, \bibinfo {author} {\bibfnamefont {S.}~\bibnamefont {Xue}}, \bibinfo {author} {\bibfnamefont {Y.}~\bibnamefont {Wang}}, \bibinfo {author} {\bibfnamefont {Y.}~\bibnamefont {Liu}}, \bibinfo {author} {\bibfnamefont {J.}~\bibnamefont {Ding}}, \bibinfo {author} {\bibfnamefont {W.}~\bibnamefont {Shi}}, \bibinfo {author} {\bibfnamefont {D.}~\bibnamefont {Wang}}, \bibinfo {author} {\bibfnamefont {Y.}~\bibnamefont {Liu}}, \bibinfo {author} {\bibfnamefont {X.}~\bibnamefont {Fu}}, \bibinfo {author} {\bibfnamefont {G.}~\bibnamefont {Huang}}, \bibinfo {author} {\bibfnamefont {A.}~\bibnamefont {Huang}}, \bibinfo {author} {\bibfnamefont {M.}~\bibnamefont {Deng}},\ and\ \bibinfo {author} {\bibfnamefont {J.}~\bibnamefont {Wu}},\ }\href@noop {} {\bibfield  {journal} {\bibinfo  {journal} {Opt. Lett.}\ }\textbf {\bibinfo {volume} {48}},\ \bibinfo {pages} {5197} (\bibinfo {year} {2023})}\BibitemShut {NoStop}%
\end{thebibliography}
\end{document}


\title{Supplemental Material} 

\setcounter{equation}{0}
\setcounter{figure}{0}
\setcounter{table}{0}
\setcounter{page}{1}
\setcounter{section}{0}
\makeatletter
\renewcommand{\theequation}{S\arabic{equation}}
\renewcommand{\thefigure}{S\arabic{figure}}
\renewcommand{\bibnumfmt}[1]{[S#1]}
\renewcommand{\citenumfont}[1]{S#1}
\newcommand{\koushik}[1]{\textcolor{blue}{#1}}



\maketitle

\section{Classical preprocessing}
\label{bitlength}

\begin{figure}[b]
    \centering
    \includegraphics[width=0.45\textwidth]{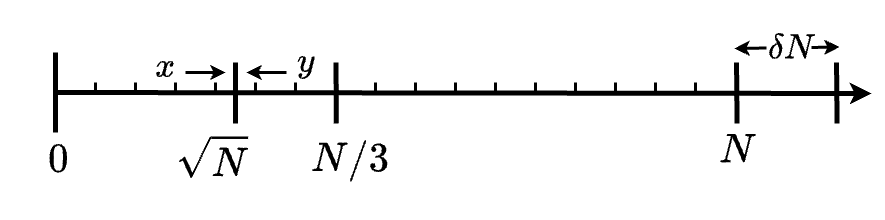}
    \caption{\textbf{Classical preprocessing for the proposed factorization.} For a given integer $N$, its factors $x$ and $y$ are less than $N/3$ and place on the either side of $\sqrt{N}$ on the number line. When $N$ is a product of two successive prime numbers, the distace between $x,y$ and $\sqrt{N}$ becomes minimal and it is possible to find a $\delta N$ so that the length of the corresponding bit-strings can be calculated according to Eq.~\eqref{bitstringlength}. }
   \label{fig:bitlength}
\end{figure}

For a general case of a factorization of a semiprime number, considering $x<y$, with both being an odd number, we can define the range as,
\begin{equation}
    3 \leq x \leq \sqrt{N}; \;\; \sqrt{N} \leq y \leq \frac{N}{3}.
   \label{range2}
\end{equation}
and the corresponding length of the strings (largest possible) can thus be upper bounded by,
\begin{equation}
     n_x = m\left(\lfloor\sqrt{N}\rfloor_{o}\right), \; n_y=m\left(\left\lfloor\frac{N}{3}\right\rfloor\right),
     \label{nxny2}
\end{equation}
where, $\lfloor a\rfloor$ ($\lfloor a\rfloor_{o}$) denotes the largest (odd) integer not larger than $a$,  while $m(b)$ denotes the smallest number of bits required for representing $b$.
However when we choose $N$s which has successive prime numbers as factors the range of $x$ and $y$ can be restricted around $\sqrt{N}$ only (see Fig.~\ref{fig:bitlength}) and the relations in Eq.~\eqref{range2} can be written as,
\begin{equation}
    x \leq \sqrt{N} \leq y.
    \label{xy}
\end{equation}
For such type of $N$s, Eq.~\eqref{nxny2} can be crudely modified to a general form as,
\begin{equation}
     n_x = m\left(\lfloor\sqrt{N}\rfloor_{o}\right) = m(\mathcal{F} -1), \; n_y = m(\mathcal{F}),
     \label{nxnymod}
\end{equation}
where $\mathcal{F} = \lceil\sqrt{N}\rceil$ and $m(b)$ denotes the smallest number of bits required for representing $b$. In most cases, this definition holds for successive prime numbers where we observe $n_x = n_y$ because the quantity $\lceil\sqrt{N}\rceil - \lfloor\sqrt{N}\rfloor$ is always equal to $1$. However this is not guaranteed, as the gap between two successive prime numbers is not constant or changes linearly \cite{AresFactor}. Therefore we define an offset $\delta N$ such that it gives the squared root distance between $N$ and $\mathcal{F}^2$ i.e., 
\begin{equation}
    \delta N = \sqrt{\mathcal{F}^2-N}.
\end{equation}
Interestingly, $\delta N^2$ is always a squared number as long as $N$ is chosen as a product of two successive prime numbers. Finally, with this offset, the upper bound on the length of the bitstrings can be defined as, 
\begin{equation}
    n_x = m ( \mathcal{F} - \delta N ); \quad n_y = m ( \mathcal{F} + \delta N ).
    \label{bitstringlength}
\end{equation}

\section{Encoding of the Factorization problem into a suitable Hamiltonian}
\label{sec:fact}

Let us consider $N$ to be the number to be factorized into two prime numbers $p$ and $q$ where $p,q \geq 3$. $N$, $p$, and $q$ are real integers and can be chosen as odd numbers without loss of generality. To find $p$ and $q$, we can design an optimization problem with an objective function, given by~\cite{peng2008quantum1},
\begin{equation}
    f(x,y) = (N-xy)^2, \;\text{where,} \; x,y \in \mathbb{Z}^+,
    \label{optimal}
\end{equation}
such that for optimal values of $x$ and $y$, $f(x,y)|_{x_c,y_c}=0$. Note that $x_c$ and $y_c$ are the bit-strings, equal to the decimal counterparts $p$ and $q$ respectively, representing the solution and the lengths of these strings are apriori unknown. However, one needs to define an approximate range of $x$ and $y$ to determine the dimension of the Hilbert space. considering, $x<y$, with both being an odd number, the range can be defined as:
The otimization problem defined by using Eq.~\eqref{optimal} can be encoded in the ground state of a quantum system, defined by the problem Hamiltonian \cite{HegadeFactor},
\begin{equation}
\left[NI- \left(\sum_{l=0}^{n_x - 1} 2^{l} \hat{x}_{l} \right)\left(\sum_{m=0}^{n_y-1} 2^{m} \hat{y}_{m} \right)\right]^2,
 \label{optHamilt}
\end{equation} 
Note that, the solution $x_c$ and $y_c$ are bitstrings of length $n_x$ and $n_y$ respectively which can be estimated as in Eq.~\eqref{bitstringlength}.
generally, Eq.~\eqref{optHamilt} represents a quantum system that requires a total ($n_x + n_y $) number of bits to describe its eigenstates in the computational basis. However, combining the observation that every odd binary number begins and ends with $1$, allows us to reduce the lengths of both $n_x$ and $n_y$. This consideration reduces the number of unknown bits for the factors $x$ and $y$ becomes $n_x-2$ and $n_y-2$ respectively, reducing the dimension of the Hilbert space to $n_x + n_y - 4$. Therefore, we replace the first and the last bit with $1$ i.e., $\hat{x}_0 = \hat{x}_{n_x - 1} = I$ and $\hat{y}_0 = \hat{y}_{n_y - 1} = I$. Furthermore, in the present study, we demonstrate an example factorization of $35 = 5 \times 7$ for which the Hamiltonian in Eq. \eqref{optHamilt} can be reduced to:
\begin{equation}
    H_P = \left[NI- \left(4 I + 2 \frac{I - \sigma_z}{2} + I\right) \otimes \left(4 I + 2\frac{I - \sigma_z}{2} + I \right)\right]^2,
    \label{optHamilt2}
\end{equation}  
where, $\hat{x}_1 = (I - \sigma_z)/2$ and $\hat{y}_1 = (I - \sigma_z)/2$, which amount to the projection on the eigenvector of $\sigma_z$ corresponding to outcome $1$. Note that, For $N = 35$, the length of the bit-string from Eq.~\eqref{nxnymod} can be calculated as $n_x = n_y = 3$ which thereby encodes the solution in the state $\ket{x_1 y_1}$.

\section{Qubit scaling}
\label{qubitscaling}
With $N$ being an $l$-bit integer,
the maximum number of bits required to represent $x$ and $y$ \eqref{nxny2} can be upper bounded by
\begin{equation}
    n_x = m\left(\lfloor\sqrt{N}\rfloor_{o}\right) \leq \left\lfloor \frac{l+1}{2}\right\rfloor, \; n_y=m\left(\left\lfloor\frac{N}{3}\right\rfloor\right) \leq l - 2.
\end{equation}
Therefore, the total number of bits, and consequently the total number of qubits, required to represent the pair $(x, y)$ satisfies the inequality:
\begin{equation}
    n_x + n_y \leq \left\lfloor \frac{l+1}{2} \right\rfloor + (l - 2) \sim \mathcal{O}\left(\frac{3l}{2}\right).
\end{equation}
By comparison, Shor's algorithm requires at least $2l$ qubits for factoring an $l$-bit number, yielding a resource scaling of $\mathcal{O}(2l)$.

In the special case where $p$ and $q$ are successive prime numbers (as in \eqref{xy}), the bit-length difference between $x$ and $y$ becomes minimal, i.e., $n_y = n_x + 1$ as $\delta N \to 0$ for large $N$. In this scenario, the total number of (qu)bits required is bounded by:
\begin{equation}
    n_x + n_y \leq 2\left\lfloor \frac{l+1}{2} \right\rfloor + 1 \sim \mathcal{O}(l),  
\end{equation}

offering a scaling advantage over a previous quantum adiabatic factoring method \cite{peng2008quantum1}, which retains its scaling $\mathcal{O}\left(\frac{3l}{2}\right)$ even in this case.

\section{Energy value phase dependence}
\label{app:energy_landscape}
\begin{figure}[t]
    \centering
\includegraphics[width=0.9\linewidth]{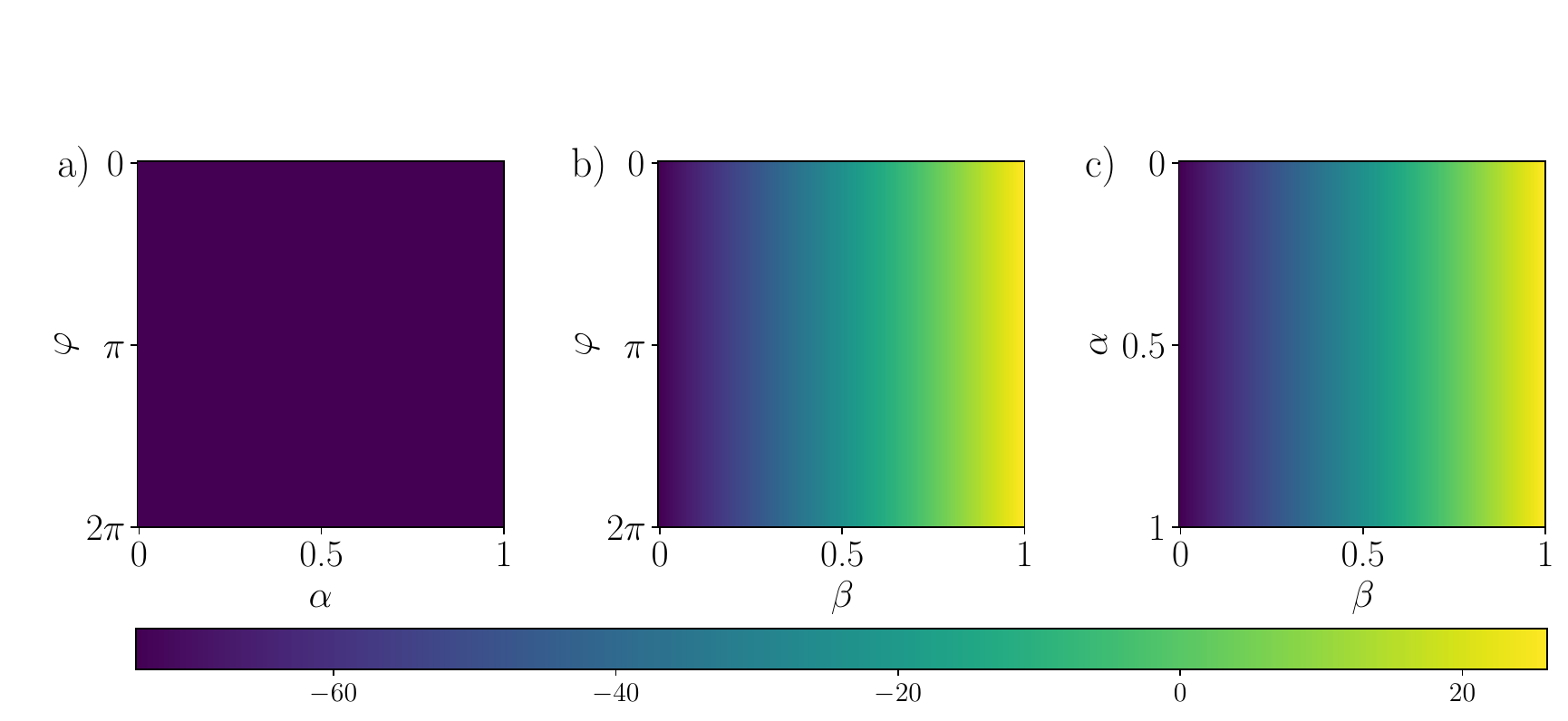}
    \caption{\textbf{Energy landscape.} The heatmaps show the expected value of the energy of differently parametrized states with respect to the Hamiltonian $H_p$ (see Eq. (4) of the main text). \textbf{a)} For the states $\sqrt{\alpha}\ket{01}+e^{-i\varphi} \sqrt{1-\alpha}\ket{10}$. All linear superpositions of the states $\ket{01}$ and $\ket{10}$ are degenerate and correspond to the ground state energy of $H_p$.
    \textbf{b)} On the state $\sqrt{\beta}\ket{00}+e^{-i\varphi} \sqrt{1-\beta}\ket{01}$. The states' energy increases monotonically in $\beta$, while all states of equal $\beta$ but different phase exhibit the same energy.
    \textbf{c)} The state $\sqrt{\beta}\ket{00} + \sqrt{1-\beta}\left(\sqrt{\alpha}\ket{01} + \sqrt{1-\alpha}\ket{10} \right)$. Again, all states with equal amplitudes of $\ket{00}$ are energy degenerate, whereas the states energy increases monotonically with $\beta$.
    }
    \label{fig:energy_landscape}
\end{figure}
 The search for the ground state of the Hamiltonian $H_p$ starts from random parameters for the twelve internal and external phase shifters ($\theta_0$ and $\phi_0$) composing our circuit. This corresponds to a random unitary operation $U(\theta_0, \phi_0)$, applied to the fixed input Fock state that we send into our circuit, i.e. $|1,0,0,0\rangle$. 
At this point, in order to proceed with the optimization, we need to estimate the corresponding energy value, which amounts to 

\begin{equation}
\label{eq:energy_unitary_app}
    \mathcal{E}(\vartheta,\varphi) = \bra{1,0,0,0}U^{\dag} (\vartheta, \varphi) H_p U (\vartheta, \varphi) \ket{1,0,0,0}
\end{equation}

In general, to this aim, we should perform a full quantum state tomography to fully reconstruct  $U(\theta_0, \phi_0)|1,0,0,0\rangle$, which will amount to 
\begin{equation}
\label{eq:energy_unitary_app1}
\sum_{S_i}\gamma_i |S_i\rangle,
\end{equation}
where $|S_i\rangle$ indicates all the permutations of one photon over four modes and $\gamma_i$ are complex values. However, due to the form of the Hamiltonian $H_p$, reported in Eq. \eqref{optHamilt2} of the main text, this operation is not necessary and just one measurement in the computational basis is sufficient. This implies that we only have access to the absolute values $|\gamma_i|$, given by the photon statistics at the four outcomes of the circuit. Indeed, as shown in Fig. \ref{fig:energy_landscape}, all of the linear combinations of $|0,1,0,0\rangle$ and $|0,0,1,0\rangle$, which are the eigenstates of the Hamiltonian corresponding to the minimum energy, are degenerate. This implies that, to reach the ground state, no knowledge of the relative phase terms among the basis elements is required, drastically reducing the number of measurements to carry out.
